\documentclass[%
 reprint,
 onecolumn,
 amsmath,amssymb,
 aps,
]{revtex4-2}

\usepackage[english]{babel}
\usepackage{graphicx}
\usepackage{dcolumn}
\usepackage{bm}
\usepackage{xcolor}
\usepackage{ragged2e}  
\usepackage{multirow}
\usepackage[frak = euler]{mathalpha}
\usepackage[citecolor=blue, colorlinks, linkcolor=blue, urlcolor=blue,]{hyperref}

\newcommand{\be}{\begin{equation}}
\newcommand{\ee}{\end{equation}}
\newcommand{\bea}{\begin{eqnarray}}
\newcommand{\eea}{\end{eqnarray}}
\newcommand{\la}{\langle}
\newcommand{\ra}{\rangle}

\begin{document}

\title{Numerical Demonstration of Kolmogorov Scaling in Magnetohydrodynamic Turbulence}

\author{Manthan Verma}
\email{manver@iitk.ac.in}
\affiliation{Department of Physics, Indian Institute of Technology, Kanpur 208016, India}

\author{Abhishek K. Jha}
\email{abhijha@iitk.ac.in}
\affiliation{Department of Physics, Indian Institute of Technology, Kanpur 208016, India}

\author{Shashwat Nirgudkar}
\email{shashwatmm@iitk.ac.in}
\affiliation{Department of Physics, Indian Institute of Technology, Kanpur 208016, India}

\author{Mahendra K. Verma}
\email{mkv@iitk.ac.in}
\affiliation{Department of Physics, Indian Institute of Technology, Kanpur 208016, India}
\affiliation{Kotak School of Sustainability, Indian Institute of Technology, Kanpur 208016, India}

\begin{abstract}
The two leading models of isotropic magnetohydrodynamic (MHD) turbulence have competing predictions: $k^{-5/3}$ (Kolmogorov) and $k^{-3/2}$ (Iroshnikov-Kraichnan) scalings. This paper identifies the valid MHD turbulence model using high-resolution numerical simulations and diagnostics—structure functions, intermittency exponents, and energy spectra and fluxes of imbalanced MHD.  The energy spectra of our forced MHD simulations on $8192^2$, $4096^2$,  $1024^3$, and $512^3$ grids support Kolmogorov's $k^{-5/3}$ spectrum over Iroshnikov-Kraichnan’s $k^{-3/2}$ spectrum, but the difference in the spectral exponents is small. However, the numerically computed third-order structure functions and intermittency exponents support Kolmogorov scaling in both two and three dimensions.  Also, the energy fluxes of the imbalanced MHD follow the predictions of Kolmogorov scaling. These results would help in a better modelling of solar wind, solar corona, and dynamos.
\end{abstract}

\maketitle

\section{Introduction}
\label{sec:intro}

Magnetohydrodynamic (MHD) turbulence is a complex phenomenon involving electrically conducting fluids---plasmas or liquid metals---that exhibit random behavior in the presence or absence of a magnetic field~\cite{Biskamp:book:MHDTurbulence,Verma:PR2004,Schekochihin:JPP2022,Zhou:PR2021,Sagaut:book,Miesch:SSR2015}. MHD turbulence is a key framework for many applications in astrophysics, liquid metals, engineering flows, and fusion reactors~{\hypersetup{citecolor=blue}\cite{Choudhuri:book:Astro,Miesch:SSR2015,Zhou:book,Zhou:ARFM2025,Zhou:PR2017_2}}. However, modeling this phenomenon remains challenging due to several nonlinear terms and fields (magnetic field \textbf{b} and velocity field \textbf{u}). Among many theories,  popular but vigorously contested, are the Iroshnikov-Kraichnan model~\cite{Iroshnikov:SA1964,Kraichnan:PF1965MHD} and the Kolmogorov-like model~\cite{Marsch:RMA1991,Goldreich:ApJ1995} for isotropic and anisotropic MHD turbulence. In this paper, we will focus on the \textit{isotropic} MHD turbulence.

Linear MHD has two kinds of Alfv\'{e}n waves that are conveniently expressed in terms of Els\"{a}sser variables, ${\bf z^\pm = u \pm b}$.   For a wavenumber \textbf{k}, ${\bf z}^+({\bf k})$ and ${\bf z}^-({\bf k})$ travel   antiparallel and parallel to  the mean magnetic field  ${\bf B}_0$, respectively.  Nonlinearity induces interactions among  ${\bf z}^+({\bf k})$ and ${\bf z}^-({\bf k})$~\cite{Biskamp:book:MHDTurbulence}.  For a strong ${\bf B}_0$, \citet{Iroshnikov:SA1964} and \citet{Kraichnan:PF1965MHD}  argued that the foward and backward waves interact for a short duration, which is the Alfv\'{e}n time scale $(k B_0)^{-1}$.  Iroshnikov and Kraichnan postulated the  following \textit{isotropic} kinetic energy spectrum [$E_u(k)$] and magnetic energy spectrum [$E_b(k)$]:
\be 
E_u(k) \approx E_b(k) \approx K_\mathrm{IK} (\epsilon^T B_0)^{1/2} k^{-3/2},
\label{eq:IK}
\ee
where $\epsilon^T$ is the total dissipation rate; and $K_\mathrm{IK}$ is a constant~\footnote{Historically, \citet{Kraichnan:JFM1959} predicted $-3/2$ spectral exponent for hydrodynamic turbulence.  Using Direct Interaction Approximation (DIA), Kraichnan argued that $E(k) \sim (\epsilon_u U_0)^{1/2} k^{-3/2}$, where $U_0$ is the rms velocity and $\epsilon_u$ is the energy dissipation rate.  However, this theory was abandoned because of disagreements with the experimental findings, as well as due to theoretical inconsistencies, such as a lack of Galilean invariance. Note, however, that Alfv\'{e}n velocity can be removed using Galilean invariance.  Hence, $k^{-3/2}$ remains a valid possibility for MHD turbulence.}.  In the absence of a mean magnetic field,   the large-scale magnetic field is treated as \textit{local mean magnetic field}.  This framework, referred to as \textit{IK scaling}, implicitly assumes that the nonlinear term is weaker than the linear term. Hence, the above model is a \textit{weak turbulence theory}. For this framework, \citet{Dobrowolny:PRL1980} showed that the inertial-range energy fluxes   [$\Pi^\pm(k)$] of  ${\bf z}^\pm$ are equal, i.e., $\Pi^+(k) = \Pi^-(k)$ irrespective of $E^+(k) / E^-(k)$ ratio, where $E^\pm(k)$ are the  energy spectra of ${\bf z}^\pm$.  

For weak ${\bf B}_0$, the nonlinear time scale $(k z_k^\mp)^{-1}$ will be the dominant time scale.  For such flows, called \textit{strong turbulence regime}, \citet{Marsch:RMA1991} proposed the following  \textit{isotropic} spectra~(also see \cite{Verma:PR2004}):
\be
E^\pm(k) = K^\pm \frac{(\epsilon^\pm)^{4/3}}{(\epsilon^\mp)^{2/3}} k^{-5/3},
\label{eq:Kolm}
\ee
where $K^\pm$ are  Kolmogorov's constant for MHD turbulence; $\epsilon^\pm$, the dissipation rates of ${\bf z}^\pm$,  equal the inertial-range energy fluxes $\Pi^\pm(k)$ in the steady state. This theory, termed as \textit{Kolmogorov scaling} \textit{for MHD turbulence}, predicts that $E^+(k)/E^-(k) = (K^+/K^-) [\Pi^+(k)/\Pi^-(k)]^2$.  Using the harmonic mean of the nonlinear time scale $(k z_k^\mp)^{-1}$ and the Alfv\'{e}n time scale $(k B_0)^{-1}$, \citet{Zhou:JGR1990model} and \citet{Zhou:RMP2004} constructed a combined model that yields dual energy spectrum: $k^{-5/3}$ at small wavenumbers and $k^{-3/2}$ at large wavenumber. \citet{Verma:PP1999} argued that the \textit{renormalized or effective mean magnetic field} scales as $k^{-1/3}$, whose insertion in Eq.~(\ref{eq:IK}) leads to $k^{-5/3}$ spectrum.

The presence of a mean magnetic field $\textbf{B}_0$ makes the flow axisymmetric about the direction of the mean magnetic field. Such flows are quantified using the energy spectrum $E(k_\perp, k_\parallel)$, where $k_\perp$ and  $k_\parallel$ are, respectively, the perpendicular and parallel components of wavenumber \textbf{k} in relation to ${\bf B}_0$~\cite{Sagaut:book,Shebalin:JPP1983}.  \textit{Ring spectrum} $E(k,\theta)$, where $\theta$ is the angle between \textbf{k} and ${\bf B}_0$, is another way to quantify anisotropy~\cite{Sagaut:book,Teaca:PRE2009,Sundar:PP2017}.  Note that for anisotropic turbulence too, we can compute $E(k)$ by averaging $E(k,\theta)$ over $\theta$ for a given $k$.  \citet{Sridhar:ApJ1994}, \citet{Goldreich:ApJ1995}, and \citet{Boldyrev:PRL2006} attempted to model such anisotropic MHD turbulence. 

For the \textit{anisotropic} \textit{strong turbulence} (with ${\bf B}_0$), where $z^\pm_k \gg B_0$,    \citet{Goldreich:ApJ1995} advocated \textit{critical balance}, $k_\parallel B_0 \approx k_\perp z^\pm_k$, to derive Kolmogorov's spectrum for the Els\"{a}sser variables [$E^\pm(k_\perp) \propto k_\perp^{-5/3}$].   However, for the weak turbulence regime ($z^\pm_k \ll B_0$), \citet{Sridhar:ApJ1994}  argued that the three-wave resonant interactions are absent in MHD, and hence IK scaling is invalid for MHD turbulence.  Therefore, they employed four-wave interactions to compute the energy spectra.  However,  \citet{Galtier:JPP2000} argued against \citet{Sridhar:ApJ1994}'s theory and showed that the three-wave interactions yield $E^\pm(k_\perp) \propto  k_\perp^{-2}$.  There are several generalizations of \citet{Goldreich:ApJ1995}'s theory. For strong anisotropic MHD turbulence with ${\bf B}_0$, \citet{Lithwick:ApJ2007} proposed scaling similar to Eq.~(\ref{eq:Kolm}), except that $k$ is replaced by $k_\perp$. \citet{Boldyrev:PRL2006} argued that  $\theta_k$, the angle between \textbf{u} and \textbf{b} fields at length scale $k^{-1}$   decreases as   $k^{-1/4}$. This preferential alignment leads to a reemergence of $E^\pm(k_\perp) \propto k_\perp^{-3/2}$  for strong turbulence with ${\bf B}_0$. 

For steady, homogeneous, and isotropic hydrodynamic turbulence, \citet{Kolmogorov:DANS1941Structure,Kolmogorov:DANS1941Dissipation}  derived a relation for the third-order structure function, which is $S^u_3(l) = -(4/5) \epsilon_u l$, where $\epsilon_u$ is the dissipation rate of the kinetic energy, and $l$ is the separation between the two points of interest (also see \citet{Obukhov:1949Scalar}). Note that the above relation is exact in the limit of vanishing viscosity, hence its verification requires a significantly large Renolds number~\cite{Sagaut:book}. \citet{Politano:GRL1998} generalized Kolmogorov's derivation to MHD turbulence with $B_0=0$ and showed that in the limit of vanishing viscosity and magnetic diffusivity, the third-order structure functions for MHD follows $S^\pm_3(l) = - (4/d) \epsilon^\pm l$, where \textit{d} is the space dimension. A straightforward dimensional analysis of $S^\pm_3(l)$ suggests that $\la {\bf z^\pm(x+l) \cdot z^\pm(x)} \ra \sim l^{2/3}$.
The Fourier transform of this relationship directly yields a Kolmogorov-like energy spectrum for isotropic MHD turbulence (with ${\bf B}_0=0$)~\cite{Biskamp:book:MHDTurbulence}.  Thus, the exact results of \citet{Politano:GRL1998} supports Kolmogorov scaling strongly. Note, however,  that generalization to structure functions other than $S^\pm_3(l)$ is more complex due to intermittency~\cite{Meneveau:PRL1987,She:PRL1993}. \citet{Grauer:PLA1994,Politano:PRE1995}, and \citet{Biskamp:PP2000} generalized \citet{She:PRL1993}'s framework  to derive intermittency exponents for MHD turbulence. We will consider these diagnostics in this paper.

In the past, researchers have attempted to validate the above models using numerical simulations and solar wind observations, some of which are summarized below. Before 2001, most numerical simulations investigated the properties of isotropic MHD turbulence. 
In particular, Biskamp, Welter, M\"{u}ller, and Schwarz~\cite{Biskamp:PFB1989,Muller:PRL2000,Biskamp:PP2001,Biskamp:PP2000} simulated two-dimensional (2D) and three-dimensional (3D) MHD turbulence on moderate grids. They showed that the energy spectra and multiscaling exponents for 3D MHD turbulence follow Kolmogorov scaling. In contrast, their numerical data for the 2D MHD turbulence support the IK spectrum,  but their multiscaling exponents are inconclusive.  However, Biskamp and coworkers consistently observed that the third-order structure functions follow Kolmogorov scaling in both 2D and 3D.   \citet{Verma:JGR1996DNS} simulated MHD turbulence on  $512^2$ and $128^3$ grids for various $E^+/E^-$ and ${\bf B}_0$, and observed agreement with  Kolmogorov scaling.
\citet{Sorriso:PP2002} and \citet{Mininni:PRE2009a} computed the third-order structure function and observed agreement with the predictions of \citet{Politano:GRL1998}. In the companion Letter~\cite{Verma:MHD_PRL}, we show that the total energy and cross helicity show $k^{-5/3}$ energy spectra. The structure functions of these quantities are also linear in $l$. However, $E_u(k) \sim k^{-3/2}$  and $E_b(k) \sim k^{-5/3}$, as is observed by several researchers in the past~\cite{Alexakis:PRL2013,Podesta:ApJ2007,Jiang:JFM2023}; we attribute the above discrepancy to the energy transfers from the magnetic field to the velocity field.

Since 2006, many numerical works have been on the anisotropic MHD turbulence. Using numerical simulations, \citet{Cho:ApJ2000} verified Goldreich-Sridhar model~\cite{Goldreich:ApJ1995}, whereas  \citet{Mason:PRL2006,Mason:PRE2008,Perez:PRL2009}, and \citet{Podesta:PP2010}  found agreement with \citet{Boldyrev:PRL2006}'s predictions, including $\theta_k \propto k^{-1/4}$.  In contrast, \citet{Beresnyak:PRL2011,Beresnyak:LR2019} reported $k_\perp^{-5/3}$ energy spectrum.   Many numerical works on the anisotropic MHD turbulence employed reduced magnetohydrodynamics (RMHD)  with ${\bf B}_0$.  Even though \citet{Galtier:JPP2000}'s calculations are analytically sound, most numerical simulations of MHD turbulence do not appear to support $k^{-2}$ energy spectrum. 

Solar wind provides an observational platform for testing MHD turbulence models.  The frequency spectrum measured along a spacecraft trajectory is translated to $E(k)$ using Taylor's frozen-in hypothesis~\cite{Taylor:PRS1938}.  Note that the solar wind is anisotropic due to the Parker field that acts as a mean field. Still, many earlier works focused on the isotropic energy spectrum $E(k)$. In particular, \citet{Matthaeus:JGR1982rugged}, and \citet{Tu:SSR1995} showed that $E_u(k), E_b(k)$, and $E^\pm(k)$  nearly follow Kolmogorov's $k^{-5/3}$ scaling.   On the other hand, \citet{Podesta:PP2010} reported that the energy spectrum of the total energy follows $k^{-3/2}$ spectrum for the large normalized cross helicity $\sigma_c = \int d{\bf r} {\bf u \cdot b}/[\int  d{\bf r} (u^2+b^2)]$, but it follows  $k^{-5/3}$ spectrum when  $\sigma_c \approx 0$. Since 2007, researchers have attempted to verify the predictions of anisotropic turbulence models.  For example, \citet{Podesta:ApJ2007} reported Kolmogorov scaling for the magnetic field and IK scaling for the velocity field. In addition, \citet{Tu:SSR1995,Podesta:PP2010},  and \citet{Parashar:PRL2018} observed that $\cos \theta_k \approx $ const.~for the inertial range $k$; these observations contradict \citet{Boldyrev:PRL2006}'s predictions that  $ \theta_k \sim k^{-1/4} $. On another front, \citet{SorrisoValvo:PRL2007} computed the third-order Structure functions using spacecraft data and reported a reasonable agreement with the predictions of \citet{Politano:GRL1998}; this result is consistent with Kolmogorov’s scaling. However, the intermittency exponents computed by \citet{Bruno:LR2013} neither follow Kolmogorov scaling nor IK scaling.

As discussed above, mean magnetic field plays a key role in MHD turbulence. The other important factors in MHD turbulence are the cross helicity, magnetic helicity, and magnetic Prandtl number (ratio of kinematic viscosity and magnetic diffusivity). In this paper, we limit our scope on the validation of IK and Kolmogorov scalings for \textit{isotropic} MHD turbulence. Also, we do not delve into other important aspects such as dynamic alignment, decay rates, etc.  We refer the reader to \citet{Biskamp:book:MHDTurbulence}, \citet{Sagaut:book}, and \citet{Miesch:SSR2015} for a broader discussion on MHD turbulence.

In this paper, we investigate isotropic MHD turbulence  using high-resolution simulations and test whether IK scaling or Kolmogorov scaling is valid. The isotropic case is applicable when the mean field is either absent or small. In addition, isotropic turbulence forms a baseline for more complex models.  Note, however, that the spectral indices ($-5/3$ and $-3/2$) of the two competing theories are too close to unambiguously distinguish them using numerical simulations. Hence, we employ three other tests to validate the models.   We compute the energy fluxes $\Pi^\pm(k)$, third-order structure functions $S_3^\pm(l)$, and the intermittency exponents.   We show that these diagnostics support Kolmogorov scaling over IK scaling.
 
The outline of this paper is as follows. In Sec. \ref{sec:Theory} we present the governing equations and turbulence models. Section \ref{sec:num_method}  contains details on our numerical method and parameters. In Sec.~\ref{sec:numerical_res} we present the numerical results on the normalized energy spectrum, fluxes, structure functions,  and intermittency exponents.  Section \ref{sec:past} contains comparisons between our work and the past results.  We conclude in Sec.~\ref{sec:Conclusions}.

\section{Governing equations, forcing scheme, and turbulence models }
\label{sec:Theory}
In this section, we describe the governing equations of MHD and those of energy fluxes. In addition, we describe the forcing scheme used in the paper and the leading MHD turbulence models.

\subsection{Equations for the energy fluxes}
The equations for incompressible MHD are~\cite{Biskamp:book:MHDTurbulence}
\bea
\frac{\partial \mathbf{u}}{\partial t} + (\mathbf{u} \cdot \mathbf{\nabla}) \mathbf{u} &=& -\nabla p + (\mathbf{B} \cdot \mathbf{\nabla}) \mathbf{B} + \mathbf{F_u} + \nu \nabla^2 \mathbf{u},
\label{eqn-1} \\
\frac{\partial \mathbf{B}}{\partial t} + (\mathbf{u} \cdot \mathbf{\nabla}) \mathbf{B} &=& (\mathbf{B} \cdot \mathbf{\nabla}) \mathbf{u} + \mathbf{F_b} + \eta \nabla^2 \mathbf{B},
\label{eqn-2} \\
\mathbf{\nabla} \cdot \mathbf{u} &=& 0,
\label{eqn-3} \\
\mathbf{\nabla} \cdot \mathbf{B} &=& 0,
\label{eqn-4}
\eea
where $\mathbf{u},\mathbf{B}, p$ are the velocity, magnetic, and pressure (magnetic+thermodynamic) fields, respectively;   $\nu$ and $\eta$ are the kinematic viscosity and magnetic diffusivity, respectively; $\mathbf{F}_u$ and $\mathbf{F}_b$ are the external forces applied to the velocity and magnetic fields, respectively. The magnetic field is separated into its mean (${\bf B}_0$) and fluctuations (${\bf b}$), i.e., ${\bf B = B}_0 + {\bf b}$. In this paper, we work with Els\"{a}sser variables, $\mathbf{z^\pm}=\mathbf{u}\pm \mathbf{b}$, because they have constant energy fluxes in the inertial range. There evolution equations of $\mathbf{z^\pm}$ are
\bea
\frac{\partial \mathbf{z^\pm}}{\partial t} 
\mp (\mathbf{B}_0 \cdot \mathbf{\nabla}) \mathbf{z^\pm}
+ (\mathbf{z^\mp} \cdot \mathbf{\nabla}) \mathbf{z^\pm} &=& -\nabla p + \mathbf{F_{z^\pm}} + \nu^{+} \nabla^2 \mathbf{z^{\pm}} + \nu^{-} \nabla^2 \mathbf{z^{\mp}},
\label{eqn-5} \\
\mathbf{\nabla} \cdot \mathbf{z^{\pm}} &=& 0,
\label{eqn-6}
\eea
where   $\mathbf{F}_{z^\pm}= \mathbf{F}_u \pm \mathbf{F}_b $ are the external forces acting on $\mathbf{z^\pm}$; and $\nu^\pm= (\nu \pm \eta )/2$. In this paper, we choose $\nu = \eta$ that makes $\nu^-=0$ and $\nu^+ = \nu = \eta$.

The nonlinear terms, $(\mathbf{z^\mp} \cdot \mathbf{\nabla}) \mathbf{z^\pm}$, yield multiscale energy transfers, which are conveniently described using Fourier transform~\cite{Biskamp:book:MHDTurbulence,Verma:PR2004,Verma:book:ET}. For ${\bf z}^\pm$ variables, the modal energy $E^\pm({\bf k}) = |{\bf z}^\pm({\bf k})|^2/2$, whose dynamical equations are~\cite{Verma:PR2004,Verma:book:ET}
\bea
\frac{d}{dt}E^\pm\mathbf{(k)} &=& \sum_{\mathbf{p}} S^\pm({\bf k|p|q})  + \sum_{\mathbf{k}} \mathrm{Real}[{\mathbf{F}_{z^\pm}({\bf k}) \cdot \mathbf{z}^{\pm*}({\bf k})}] - 2  k^2 \nu^+ E^\pm \mathbf{(k)} \\
&=& T^{\pm}\mathbf{(k)} + \mathcal{F}^{\pm}(\mathbf{k}) - D^{\pm}(\mathbf{k}),
\label{eq:Ek_dot}
\eea
where  $\mathcal{F}^{\pm}(\bold{k})$ are energy injection rates to  $\mathbf{z}^\pm({\bf k})$ by the external forces $\mathbf{F}_{z^\pm}$;  $D^{\pm}(\bold{k})$ are the dissipation rates of  $\mathbf{z}^\pm(\bold{k})$;  $T^{\pm}\mathbf{(k)}$ are the modal energy transfers to mode $\mathbf{z}^\pm(\bold{k})$ through nonlinear interactions; and
\be 
S^\pm({\bf k|p|q}) = \mathrm{Im}[\{\mathbf{k \cdot z^\mp(q)\}\{z^\pm(p)\cdot z^{\pm*}(k)\}}]
\ee
are the \textit{mode-to-mode energy transfer rates} from ${\bf z^\pm(p)}$
to ${\bf z^\pm(k)}$ with the mediation of ${\bf z^\mp(q)}$~\cite{Dar:PD2001,Verma:PR2004}. It has been shown that there is no energy transfers between ${\bf z}^+$ and  ${\bf z}^-$~\cite{Verma:JPA2022}.

Using $T^{\pm}\mathbf{(k)}$ and $S^\pm({\bf k|p|q})$, we can define energy fluxes, $\Pi^\pm(R)$, which are the net energy transfers from the modes ${\bf z}^\pm$ within the wavenumber sphere of radius of $R$ to the modes ${\bf z}^\pm$ outside the sphere. These fluxes can be computed efficiently using~\cite{Verma:book:ET}
\bea
\Pi^{\pm}(R) = - \int_{|{\bf k}|<R} T^{\pm}({\bf k}) d{\bf k} =  - \int_{0}^{R} T^{\pm}(k) dk,
\label{eq:flux_def}
\eea
where $T^\pm(k)$ are the shell spectra related to $T^\pm({\bf k})$. For large-scale forcing, the inertial-range $\Pi^{\pm}(R)$ are constant  because the cross transfers between ${\bf z}^+$ and  ${\bf z}^-$ are absent~\cite{Verma:PR2004,Verma:JPA2022}.   Under steady state, the inertial-range energy fluxes,  the corresponding dissipation rates, and the energy injection rates are all equal statistically~\cite{Lesieur:book:Turbulence,Verma:book:ET}. That is,
\be
\la \Pi^\pm(R) \ra = \epsilon^\pm = 2\nu \int_0^\infty k^2 E^\pm(k) dk  \approx \epsilon^\pm_\mathrm{inj},
\label{eq:epsilon_integral}
\ee 
when $R$ in the inertial range.

MHD turbulence has several other energy fluxes, e.g., $\Pi^{u<}_{u>}(R)$, $\Pi^{b<}_{b>}(R)$, $\Pi^{u<}_{b>}(R)$, $\Pi^{b<}_{u>}(R)$, where $\Pi^{X<}_{Y>}(R)$ is the net energy transfer from the field ${\bf X}$ within the wavenumber sphere of radius of $R$ to the field ${\bf Y}$ outside the sphere~\cite{Dar:PD2001,Verma:PR2004}. The interactions between \textbf{u} and \textbf{b} facilitate energy exchange between \textbf{u} and \textbf{b}. As a result,  we do not have constant energy fluxes for \textbf{u} and \textbf{b}, which leads deviations of $E_u(k)$ and $E_b(k)$  from the $k^{-5/3}$ spectrum~\cite{Verma:JPA2022,Verma:Fluid2021}.  For example, a steeper $E_u(k)~(\sim k^{-0.73})$ than $E_b(k)$ ($ \sim k^{-0.68}$) reported by \citet{Verma:Fluid2021} can be attributed to the energy transfers from \textbf{u} to \textbf{b}. In contrast, the inertial-range $\Pi^\pm(k)$ are constant because of an absence of cross transfers between ${\bf z}^+$ and ${\bf z}^-$. This is the reason why we focus on $\Pi^\pm(k)$ and $E^\pm(k)$ in this paper.

\subsection{Our forcing algorithm}

In this paper, we focus on the statistical properties of steady MHD turbulence. According to \textit{antidynamo theorem}~\cite{Cowling:book}, in 2D MHD, the magnetic field vanishes when only the velocity field is forced. Therefore,  we force both \textbf{u} and \textbf{b} fields (or ${\bf z}^\pm$) to obtain a steady state in 2D MHD. That is, we set ${\bf F}_b \ne 0$, or we force both ${\bf z}^+$ and ${\bf z}^-$ fields in 2D simulations, as well as in 3D simulations for uniformity. Following Kolmogorov theory of turbulence, we force only the large-scale modes to obtain constant $\Pi^{\pm}(R)$ in the inertial range. 

There are several ways to force a turbulence flow. In this paper, we adopt the schemes proposed by \citet{Carati:JoT2006} and \citet{Teimurazov:JAMTP2018}, and force a wavenumber band in which $\epsilon^+$ and $\epsilon^-$ are divided equally among all the modes of the shell. For convenience,  the variables in this subsection are represented using tilde, that is, $\tilde{f} = f({\bf k})$ (modal $f$). With this, the force $\mathbf{\Tilde{F}_{z^\pm}}$ are
 \bea
\mathbf{\Tilde{F}_{z^+}} &=& \mathbf{\Tilde{\alpha} \Tilde{z}^+ + \Tilde{\beta} \Tilde{z}^-}, 
\label{eq:Fp}\\
\mathbf{\Tilde{F}_{z^-}} &=& \mathbf{\Tilde{\gamma} \Tilde{z}^- + \Tilde{\beta} \Tilde{z}^+}, 
\label{eq:Fm}
\eea
where $\tilde{\alpha},\tilde{\beta}$, and $\tilde{\gamma}$ are constants for the wavenumber \textbf{k}. We multiply Eq.~(\ref{eq:Fp}) and (\ref{eq:Fm}) 
 with $\mathbf{\Tilde{z}^{+*}}$ and $\mathbf{\Tilde{z}^{-*}}$, respectively, and then take their respective real parts, which yields
\bea
\tilde{\epsilon}_{\mathrm{inj}}^+ &=& 2(\Tilde{\alpha} \Tilde{E}^+ + \Tilde{\beta} \Tilde{E}_R), 
\label{eq:Fplus}\\ 
\tilde{\epsilon}_{\mathrm{inj}}^- &=& 2(\Tilde{\gamma} \Tilde{E}^- + \Tilde{\beta} \Tilde{E}_R),
\label{eq:Fminus}
\eea
where $\Tilde{\epsilon}_{\mathrm{inj}}^\pm$ are the modal injection rates, and $\Tilde{E}_R = \Tilde{E}_u - \Tilde{E}_b$ is the modal \textit{reduced energy}~\cite{Biskamp:book:MHDTurbulence}.

So far, we have two equations [Eqs.~(\ref{eq:Fplus}, \ref{eq:Fminus})] and three unknowns, which are $\tilde{\alpha},\tilde{\beta}$, $\tilde{\gamma}$. We invoke the conservation laws of MHD to derive the third equation. The total energies of ${\bf z}^\pm$, $E^\pm$, are conserved in dissipationless MHD (with $\nu = \eta =0$). Equations~(\ref{eq:Fplus}, \ref{eq:Fminus}) correspond to the conservation of $E^\pm$, respectively.  In addition, magnetic helicity ($H_M = {\bf A \cdot B}/2$) is conserved in 3D, and mean-square vector potential ($A^2$) is conserved in 2D, where \textbf{A} is the vector potential. Therefore, we construct the additional equation using the injection rates of these quantities ($H_M$ or $A^2$).

For 2D, using Eqs.~(\ref{eq:Fp}) and (\ref{eq:Fm}), we derive the equation for $A^2$ injection rate ($\Tilde{\epsilon}_{\mathrm{inj},A}$) as
\bea
\Tilde{\epsilon}_{\mathrm{inj},A} &=& \frac{(\Tilde{\alpha} -\Tilde{\beta}) \Tilde{E}^+ - (\Tilde{\beta} - \Tilde{\gamma}) \Tilde{E}^- + \left[ 2\Tilde{\beta} -(\Tilde{\alpha} + \Tilde{\gamma}) \right] \Tilde{E}_R}{2k^2}.
\label{eq:epsA2}
 \eea
Now, Eqs.~(\ref{eq:Fplus}, \ref{eq:Fminus}, \ref{eq:epsA2}) can be written in the matrix form as
\bea
\begin{bmatrix}
\Tilde{E}^+ & \Tilde{E}_R & 0 \\
0 & \Tilde{E}_R & \Tilde{E}^- \\
(\Tilde{E}^+-\Tilde{E}_R) & (2\Tilde{E}_R-\Tilde{E}^+-\Tilde{E}^-) & (\Tilde{E}^--\Tilde{E}_R)
\end{bmatrix}
\begin{bmatrix}
\Tilde{\alpha}\\
\Tilde{\beta} \\
\Tilde{\gamma}
\end{bmatrix}
&=&
\begin{bmatrix}
\Tilde{\epsilon}^+_\mathrm{inj}/2 \\
\Tilde{\epsilon}^-_\mathrm{inj}/2 \\
2k^2\Tilde{\epsilon}_{\mathrm{inj},A},
\end{bmatrix}
\eea
whose solution is
 \bea
 \Tilde{\alpha} &=& \frac{1}{D_1} [4\Tilde{E}^-\Tilde{E}_Rk^2 \Tilde{\epsilon}_{\mathrm{inj},A} + \Tilde{\epsilon}^+_\mathrm{inj}\Tilde{E}^-(\Tilde{E}^+ + \Tilde{E}^-) - \Tilde{E}^-\Tilde{E}_R(\Tilde{\epsilon}^+_\mathrm{inj} + \Tilde{\epsilon}^-_\mathrm{inj}) + \Tilde{E}_R^2 (\Tilde{\epsilon}^-_\mathrm{inj} - \Tilde{\epsilon}^+_\mathrm{inj})],
 \label{eqn:alpha_2d}\\
 \Tilde{\beta} &=& -\frac{1}{D_1}[4\Tilde{E}^-\Tilde{E}^+k^2 \Tilde{\epsilon}_{\mathrm{inj},A} - \Tilde{E}^+\Tilde{E}^-(\Tilde{\epsilon}^+_\mathrm{inj} + \Tilde{\epsilon}^-_\mathrm{inj}) + \Tilde{E}_R(\Tilde{E}^-\Tilde{\epsilon}^+_\mathrm{inj} + \Tilde{E}^+\Tilde{\epsilon}^-_\mathrm{inj})],
 \label{eqn:beta_2d}\\
  \Tilde{\gamma} &=& \frac{1}{D_1}[4\Tilde{E}^+\Tilde{E}_Rk^2 \Tilde{\epsilon}_{\mathrm{inj},A} + \Tilde{E}^+\Tilde{\epsilon}^-_\mathrm{inj}(\Tilde{E}^+ + \Tilde{E}^-) - \Tilde{E}^+\Tilde{E}_R(\Tilde{\epsilon}^+_\mathrm{inj} + \Tilde{\epsilon}^-_\mathrm{inj}) + \Tilde{E}_R^2(\Tilde{\epsilon}^+_\mathrm{inj} - \Tilde{\epsilon}^-_\mathrm{inj})],
 \label{eqn:gamma_2d}
 \eea
where 
\be
D_1 = 2(\Tilde{E}^+ + \Tilde{E}^-)(\Tilde{E}^+ \Tilde{E}^- - \Tilde{E}_R^2).\\
\ee

Similarly, for 3D, we derive an equation for $H_M$ injection rate ($\Tilde{\epsilon}_{\mathrm{inj},H_M}$), which is given below:
\bea
k^2\Tilde{\epsilon}_{\mathrm{inj},H_M} &=& k^2(\Tilde{\alpha} + \Tilde{\gamma} - 2\Tilde{\beta})\Tilde{H}_M + (\Tilde{\alpha} - \Tilde{\gamma})\Tilde{H}_1,
\label{eqn:magnetic_helicity_injection}\\
\Tilde{H}_1 &=& \frac{1}{2}\Im[\mathbf{k}.(\Tilde{\mathbf{b}}^* \times \Tilde{\mathbf{u}})]
\label{eqn:H1_in_magnetic_helicity_inj}.
\eea
Using same  method as described above, we obtain  $\Tilde{\alpha},\Tilde{\beta}$, and $\Tilde{\gamma}$ for 3D as
\bea
\Tilde{\alpha} &=& \frac{1}{D_2}[2k^2\Tilde{E}^-\Tilde{E}_R\Tilde{\epsilon}_{\mathrm{inj},H_M} + k^2\Tilde{H}_M(2\Tilde{E}^-\Tilde{\epsilon}^+_{\mathrm{inj}} + \Tilde{E}_R(\Tilde{\epsilon}^+_{\mathrm{inj}} - \Tilde{\epsilon}^-_{\mathrm{inj}})) + \Tilde{E}_R\Tilde{H}_1(\Tilde{\epsilon}^-_{\mathrm{inj}} - \Tilde{\epsilon}^+_{\mathrm{inj}})],
\label{eqn:alpha_3d}\\
\Tilde{\beta} &=& -\frac{1}{D_2}[2k^2\Tilde{E}^+\Tilde{E}^-\Tilde{\epsilon}_{\mathrm{inj},H_M} - k^2\Tilde{H}_M(\Tilde{E}^+\Tilde{\epsilon}^-_{\mathrm{inj}} + \Tilde{E}^-\Tilde{\epsilon}^+_{\mathrm{inj}}) + \Tilde{H}_1(\Tilde{E}^+\Tilde{\epsilon}^-_{\mathrm{inj}} - \Tilde{E}^-\Tilde{\epsilon}^+_{\mathrm{inj}})],
\label{eqn:beta_3d}\\
\Tilde{\gamma} &=& \frac{1}{D_2}[2k^2\Tilde{E}^+\Tilde{E}_R\Tilde{\epsilon}_{\mathrm{inj},H_M} + k^2\Tilde{H}_M(2\Tilde{E}^+\Tilde{\epsilon}^-_{\mathrm{inj}} + \Tilde{E}_R(\Tilde{\epsilon}^-_{\mathrm{inj}} - \Tilde{\epsilon}^+_{\mathrm{inj}})) + \Tilde{E}_R\Tilde{H}_1(\Tilde{\epsilon}^-_{\mathrm{inj}} - \Tilde{\epsilon}^+_{\mathrm{inj}})],
\label{eqn:gamma_3d}
\eea
where 
\be
D_2 = 2(k^2\Tilde{H}_M(2\Tilde{E}^+\Tilde{E}^- + \Tilde{E}_R(\Tilde{E}^- + \Tilde{E}^+)) + \Tilde{E}_R\Tilde{H}_1(\Tilde{E}^- - \Tilde{E}^+)).
\ee
The above coefficients enable appropriate energy injection in our simulations, leading to the desired statistical steady flows.

As in Kolmogorov's theory of hydrodynamic turbulence~\cite{Kolmogorov:DANS1941Dissipation,Kolmogorov:DANS1941Structure}, we expect the MHD turbulence properties to be universal under forcing at large scales.  That is, another set of random large-scale forcing with the same injection parameters should yield similar energy spectra and fluxes.  Note, however, that 2D MHD turbulence simulations are quite sensitive to initial conditions~\cite {Dar:PP1998}.   In addition, the inverse cascade of mean-square magnetic vector potential ($A^2$) leads to severe instabilities.  Hence, we adjust our 2D parameters carefully to obtain a steady state.

\subsection{Leading spectral theories of MHD turbulence}

There are several spectral theories for MHD turbulence. In this subsection we focus on Iroshnikov-Kraichnan (IK)~\cite{Iroshnikov:SA1964,Kraichnan:PF1965MHD} and Kolmogorov-like phenomenologies  \cite{Marsch:RMA1991,Goldreich:ApJ1995} that have contradictory predictions.  Note that these theories have subtle differences in the presence and absence of ${\bf B}_0$.  Refer to \citet{Biskamp:book:MHDTurbulence,Verma:PR2004,Verma:book:ET},  \citet{Schekochihin:JPP2022}, and \citet{Zhou:PR2021} for details.

Iroshnikov-Kraichnan's $k^{-3/2}$ theory is inspired by short-time interaction between the Alfv\'{e}n waves when the mean magnetic field ${\bf B}_0$ is strong.   \citet{Kraichnan:PF1965MHD} and \citet{Iroshnikov:SA1964} argued that these oppositely-moving  Alfv\'{e}n wave interact with the Alfv\'{e}n time scale, $(k B_0)^{-1}$.  Using these arguments and dimensional analysis, \citet{Kraichnan:PF1965MHD} and \citet{Iroshnikov:SA1964} derived Eq.~(\ref{eq:IK}) for the kinetic and magnetic spectra. For the isotropic case, $B_0$ is the magnitude of the large-scale magnetic field, whereas for the anisotropic case,  $B_0 = |{\bf B}_0|$  and $k \rightarrow k_\perp$ in Eq.~(\ref{eq:IK}). The above power law is referred to as \textit{IK scaling}. Further, for isotropic regime of the IK framework, \citet{Dobrowolny:PRL1980} showed that the inertial-range energy fluxes follow
\be
\Pi^+(k) = \Pi^-(k)
\label{eq:flux_equal}
\ee
irrespective of $E^+(k)/E^-(k)$ ratio. For the anisotropic framework with $z^\pm_k \ll B_0$, \citet{Sridhar:ApJ1994} showed that the three-wave resonant interactions are absent in MHD, leading to absence of IK scaling. However,  \citet{Galtier:JPP2000} showed that three-wave interactions exist in MHD turbulence, but these interactions yield  $E^\pm(k_\perp) \sim k_\perp^{-2}$.  In the IK model, the linear terms (${\bf B}_0 \cdot \nabla) {\bf z}^\pm$ dominate the nonlinear terms $({\bf z ^\mp \cdot  \nabla) z^\pm}$~\cite{Kraichnan:PF1965MHD,Biskamp:book:MHDTurbulence,Verma:PR2004,Schekochihin:JPP2022}. Hence, the IK model comes under a broad framework of \textit{weak turbulence}. 
 
When the nonlinear term dominates the linear term, turbulence is said to be \textit{strong}.  In an early work, \citet{Marsch:RMA1991} proposed  Eq.~(\ref{eq:Kolm}), which is \textit{Kolmogorov  scaling} for  the \textit{isotropic MHD turbulence}.    Note that the dissipation rates  $\epsilon^\pm$ equal the inertial-range fluxes $\Pi^\pm(k)$ in the steady state.   In this framework,  Eq.~(\ref{eq:Kolm}) leads to the following relations in the inertial range:
\be
\frac{\Pi^+(k)}{\Pi^-(k)}  =
\frac{\epsilon^+}{\epsilon^-} = \left[ \frac{E^+(k)}{E^-(k)}  \frac{K^-}{K^+} \right]^{1/2}
\ee
or
\be
\frac{E^+(k)}{E^-(k)}  =\frac{K^+}{K^-} \left[  \frac{\Pi^+(k)}{\Pi^-(k)}  \right]^2
\label{eq:Kolm_ratio}
\ee
with $K^\pm = O(1)$.  Hence, $\Pi^+(k) > \Pi^-(k)$ when $E^+(k) > E^-(k)$ and vice versa, which is contrary to  IK and \citet{Dobrowolny:PRL1980} predictions that  $\Pi^+(k) = \Pi^-(k)$ irrespective of $E^+(k) / E^-(k)$ ratio. The flows with  $E^+(k)  \ne E^-(k)$ are referred to as \textit{Alfv\'{e}nic} or \textit{imbalanced MHD}~\cite{Goldstein:ARAA1995,Lithwick:ApJ2007,Beresnyak:ApJ2008a}. The cross helicity, $\int d{\bf r} ({\bf u \cdot b})/2$, and normalized cross helicity, $\sigma_c = (E^+-E^-)/(E^++E^-) = \int d{\bf r} 2({\bf u \cdot b})/\int d{\bf r} (u^2+b^2)$ are employed as a measure of Alfv\'{e}nicity~\cite{Goldstein:ARAA1995}.

\citet{Zhou:JGR1990model} and \citet{Zhou:RMP2004} argued that the effective time scale for MHD turbulence is the harmonic mean of the nonlinear time scale $(k z_k^\mp)^{-1}$ and the Alfv\'{e}n time scale $(k B_0)^{-1}$, i.e.,
\be
\frac{1}{\tau^\pm_\mathrm{eff}} =  k z_k^\mp + k B_0.
\ee
Using dimensional analysis, they derived a combined model that yields dual energy spectra: $k^{-5/3}$ at small wavenumbers and $k^{-3/2}$ at large wavenumber. In another development, \citet{Verma:PP1999} argued that the effective mean magnetic field has the following $k$-dependence, $B_0(k) \sim (\epsilon^T)^{1/3} k^{-1/3}$, substitution of which in Eq.~(\ref{eq:IK}) yields Kolmogorov scaling.

For the strong turbulence regime of anisotropic MHD  (with ${\bf B}_0$), \citet{Goldreich:ApJ1995} argued that $k_\parallel B_0 \approx k_\perp z^\pm$, called \textit{critical balance}, and that 
\be
E(k_\perp) \sim k_\perp^{-5/3},
\ee
which is a variant of Eq.~(\ref{eq:Kolm}) for the anisotropic case. \citet{Boldyrev:PRL2006} revised Goldreich-Sridhar theory by incorporating dynamic alignment and argued that the velocity and magnetic fields are preferentially aligned, with their angular divergence $\theta_k \sim k^{-1/4}$ that leads to IK scaling ($k_\perp^{-3/2}$ spectrum). Thus,  \citet{Boldyrev:PRL2006}'s theory takes one back to the IK spectrum in the strong turbulence regime. Also see Podesta and Bhattacharjee~\cite{Podesta:AJ2010}. A cautionary remark, our paper deals with isotropic MHD turbulence (with $|{\bf B}_0| =0$), which may have different properties than those of Goldreich-Sridhar and  Boldyrev models.

In the past, researchers~\cite{Biskamp:PFB1989,	Verma:JGR1996DNS,Biskamp:PRL1999,Biskamp:PP2000,Mason:PRL2006,Beresnyak:PRL2011,Sahoo:NJP2011} performed numerical simulations and computed $E(k)$.   Unfortunately, some works support $k^{-5/3}$ spectrum, and others support $k^{-3/2}$ spectrum. Hence, these works are inconclusive because the spectral indices $-5/3$ and $-3/2$  are quite close.  Our strategy is different. Even though our high-resolution 2D simulations   support $k^{-5/3}$ reasonably well, our tests on $\Pi^\pm(k)$ are  conclusive. In Sec.~\ref{sec:energy_spec_flux}, we show that the numerical $\Pi^\pm(k)$'s are closer to Eq.~(\ref{eq:Kolm_ratio}) with $K^+ \ne K^-$ than Eq.~(\ref{eq:flux_equal}), thus favoring Kolmogorov scaling for MHD turbulence.

We employ two more tests based on structure functions. For steady, homogeneous, and isotropic hydrodynamic turbulence in 3D, under the limit of kinematic viscosity $\nu \rightarrow 0$, Kolmogorov~\cite{Kolmogorov:DANS1941Dissipation,Kolmogorov:DANS1941Structure,Obukhov:1949Scalar} showed that the third-order structure function 
\bea
S_3^u(l) =  \langle\lvert \Delta \mathbf{u}\rvert^2(\Delta \mathbf{u\cdot \hat{l})} \rangle &=& - \frac{4}{5} \epsilon_u l,
  \label{eq:S3_HD}
\eea
where $ \Delta \mathbf{u} = \mathbf{u(x+l) - u(x)} $, and $\epsilon_u$  is the dissipation rate of the kinetic energy~\cite{Frisch:book}. \citet{Politano:GRL1998} generalized Kolmogorov's structure-function derivation to  isotropic MHD turbulence (${\bf B}_0 = 0$) under the limit $\nu \rightarrow 0$ and $\eta \rightarrow 0$ and obtained the following third-order structure functions  for the Els\"{a}sser variables:
\bea
S^\pm_3(l) =  \langle\lvert \Delta \mathbf{z^\pm}\rvert^2(\Delta \mathbf{z^{\mp}\cdot \hat{l})} \rangle &=& - \frac{4}{d} \epsilon^\pm l,
  \label{eq:S3}
\eea
where $ \Delta \mathbf{z^\pm} = \mathbf{z^\pm(x+l) - z^\pm(x)} $, and $d$ is the space dimension. Equation~(\ref{eq:S3}) is an exact relation for MHD turbulence. For a simple fractal model~\cite{Frisch:book}, using Eq.~(\ref{eq:S3}) we can estimate that $\la (\Delta {\bf z})^2 \ra \sim l^{2/3}$ or $\la {\bf z(x+l) \cdot z(x)} \ra \sim C_1 - C_2 l^{2/3}$, where $C_1, C_2$ are constants.  The above relation leads to $E^\pm(k) \sim k^{-5/3}$. Our numerical results, discussed in Sec.~\ref{sec:numerical_res} validate Eq.~(\ref{eq:S3}), thus favoring Kolmogorov scaling for MHD turbulence. \citet{Politano:GRL1998} have also derived structure functions for the total energy and cross helicity, which is discussed in the companion Letter~\cite{Verma:MHD_PRL}.

\citet{Politano:PRE1995} and \citet{Biskamp:PP2000} extended the  intermittency model of \citet{She:PRL1994}  to MHD turbulence and argued that
\be
S^\pm_p(l) = \la\lvert \Delta \mathbf{z^\pm}\rvert^p \ra \sim l^{\zeta_p^\pm},
\label{eq:Sp}
\ee
where $\zeta^\pm_p$ are the \textit{intermittency exponents}.  An extension of $k^{-5/3}$ energy spectrum to intermittency analysis yields~\cite{Politano:PRE1995,Biskamp:PP2000} 
\be
 \zeta_p^\pm  =   \zeta_p^\mathrm{Kolm} =  \frac{p}{9} + 1 - \frac{1}{3^{p/3}}.
\label{eq:Sp_Kolm}
\ee
Note that $\zeta_p^+ = \zeta_p^-$ in this framework.
In contrast, for $k^{-3/2}$ energy spectrum, dimensional analysis yields~\cite{Grauer:PLA1994,Politano:PRE1995,Biskamp:PP2000} 
\be
 \zeta_p^\pm = \zeta_p^\mathrm{IK} = \frac{p}{8} + 1 - \frac{1}{2^{p/4}}.
\label{eq:Sp_IK}
\ee
Note that Eq.~(\ref{eq:S3}) is derived from the first principle, but Eqs.~(\ref{eq:Sp_Kolm},\ref{eq:Sp_IK}) are  phenomenological. For a simple fractal model~\cite{Frisch:book}, $\zeta_4^\mathrm{IK} = 1$ implies that $\la (\Delta {\bf z})^2 \ra \sim l^{1/2}$ or $\la {\bf z(x+l) \cdot z(x)} \ra \sim D_1 - D_2 l^{1/2}$, where $D_1, D_2$ are constants.  The above relation leads to $E^\pm(k) \sim k^{-3/2}$.  Note, however, that the spectral exponents get corrections due to intermittency ($\zeta^\pm_2 \ne 2/3$ or 1/2). In Sec.~\ref{sec:numerical_res} we report $S_3^\pm(l)$ and intermittency exponents computed using numerical data. Our results are closer to Eq.~(\ref{eq:Sp_Kolm}) than Eq.~(\ref{eq:Sp_IK}), thus supporting Kolmogorov scaling for MHD. However, we remark that a moderate scaling range for the structure functions requires significantly large grid resolution~\cite{Sagaut:book}. This issue also impacts extended self similarity (ESS) employed in the paper.

With this, we conclude our brief on the models for IK and Kolmogorov scalings. In Sec.~\ref{sec:numerical_res}, we contrast the two competing spectral theories using structure functions, intermittency exponents,   energy spectra, and energy fluxes as diagnostics.

\section{Numerical Method and Parameters}
\label{sec:num_method}

We numerically solve Eqs.~(\ref{eqn-5}, \ref{eqn-6})  on a $(2\pi)^d$ box ($d$ = space dimension) using \textit{Aithon}, a pseudospectral CUDA code. Accurate determination of spectral exponents and energy fluxes requires high-resolution simulations, which are computationally expensive in three dimensions. Interestingly,  \textit{absolute equilibrium theory} for MHD turbulence~\cite{Frisch:JFM1975,Kraichnan:ROPP1980,Biskamp:book:MHDTurbulence} predicts similar spectra and fluxes for the total energy in both 2D and 3D. Therefore, we employ reduced-cost 2D MHD turbulence simulations, along with 3D simulations, to investigate the scaling of MHD turbulence~\cite{Biskamp:PFB1989,Verma:JGR1996DNS}. We employ $4096^2$ and $8192^2$ grids for 2D simulations, and $512^3$ and $1024^3$ grids for 3D simulations. To maintain a steady state for all our runs, we force both ${\bf z}^+$ and ${\bf z}^-$ in the wavenumber band  $(2, 3)$  using the scheme outlined in Sec.~\ref{sec:Theory}. The forcing of the magnetic field overcomes the antidynamo theorem~\cite{Cowling:book}.

As shown in Table~\ref{tab:simulation_details}, $\epsilon^-_{\mathrm{inj}} = 0.1$ for all  the runs, but $\epsilon^+_{\mathrm{inj}}$ is chosen as 0.1, 0.3, and 0.5. For $\epsilon^+_{\mathrm{inj}}/\epsilon^-_{\mathrm{inj}} = 3$ and 5, this strategy yields $\Pi^+(k) \ne \Pi^-(k)$ and $E^+(k)/E^-(k) \ne 1$ that enables us to contrast the IK and Kolmogorov scalings. The choice of forcing band at $ (2, 3)$ (large scales) helps us obtain constant energy fluxes in the inertial range. The injection rate of magnetic helicity ($\epsilon_\mathrm{inj}^{H_M}$) in 3D is zero, but that of mean-square vector potential  ($\epsilon_\mathrm{inj}^{A}$) in 2D is small, which is required for gaining a steady state in 2D.   We also compute the structure functions for all the runs and test the scaling.

The model tests via the energy fluxes and structure functions do not require extreme resolutions.  As we show in this paper, we can reach definitive conclusions using  the aforementioned simulations with moderate resolutions in 3D and high resolutions in 2D. We employ fourth-order Runge-Kutta ($RK4$) time integration scheme, with a time stepping determined by the Courant–Friedrichs–Lewy (CFL) condition using a Courant number of $0.1$. One of the smallest timesteps $\Delta t \approx 2 \times 10^{-6}$ is for  $8192^2$ grid with $\epsilon^+_{\mathrm{inj}}/\epsilon^-_{\mathrm{inj}} = 5$.  All our simulations are well resolved with $k_\mathrm{max} {l_d}$ ranging from 3.2 to 4.0, where  $l_\mathrm{d} \approx (\nu^3/\epsilon_{\mathrm{inj}})^{1/4}$ is the Kolmogorov length, and $k_\mathrm{max} = \pi/N$ with $N$ as the grid resolution.

We choose $\nu=\eta$ for all our runs with minimum $\nu = 4.55\times 10^{-5}$ for 2D and minimum $\nu = 5.6\times 10^{-4}$ for 3D.  The corresponding Reynolds numbers Re 
 ($UL/\nu$) are $96660$ and $7055$ respectively, where $U$ is the large-scale velocity, and $L$ is the box size. The magnetic Reynolds number ($UL/\eta$) matches Re because $\nu = \eta$. See Table~\ref{tab:simulation_details} for details.  We observe that these Reynolds numbers are reasonably large to distinguish Kolmogorov and IK scaling. However, Re for the 3D runs are insufficient to study locality in MHD turbulence~\cite{Zhou:PP2007,Zhou:PP2011}.  In addition, we employ hypo-viscosity and hypo-diffusivity in all the 2D runs to minimize the inverse cascade of magnetic vector potential that increases monotonically~\cite{Pouquet:JFM1978,Kraichnan:ROPP1980}. Our hypo-dissipation terms are of the form $\nu_\mathrm{hypo} k^{-2} \mathbf{z^{\pm}}(k, t)$ with  $\nu_\mathrm{hypo} \approx 0.01$.  We also compute the hydrodynamic entropy $S_H = - \sum_k p_k \log_2 p_k$, where $p_k = E(k)/E$ with $E=(E^++E^-)/2$ as the total energy and $E(k)$ as the shell spectrum~\cite{Verma:PRF2022,Verma:PRE2024}. The entropies for all the runs are listed in Table~\ref{tab:simulation_details}. We observe that $S_H$ decreases with the increase of $\epsilon^+_\mathrm{inj}$, which indicates an increase in hydrodynamic order with the increase of $\sigma_c$. 

We performed our simulations mostly on \emph{Frontier} of the Oak Ridge Leadership Computing Facility that has total $9408$ GPU compute nodes, each containing $4$ AMD MI250X GPU cards. Some of the simulations were performed on a GPU node of Kotak School of Sustainability, IIT Kanpur. All our simulations were performed using our CUDA-based pseudo-spectral solver \emph{Aithon} that scales well up to $4096$ nodes of Frontier. Our runs took significant computing time. For example, each of the $1024^3$ runs took approximately 18 hours on $64$ nodes of Frontier.  In addition, we employ a GPU-enabled structure function  solver that speeds up our computations by many folds compared to CPU-based solvers; this solver adopts the parallelization strategy used by \citet{Sadhukhan:JOSS2021}.

\begin{table*}
\renewcommand{\tablename}{\scriptsize TABLE}
\caption{\scriptsize Simulation parameters. Runs 1 to 6 are for 2D, and Runs 7 to 12 are for 3D. The table lists grid size, kinematic viscosity ($\nu$), Re, energy injection rate to ${\bf z}^+$ ($\epsilon^+_\mathrm{inj}$), magnetic helicity injection rate ($ {\epsilon_\mathrm{inj}^{H_M}}$) for 3D, injection rate for the mean-square vector potential ($ {\epsilon_\mathrm{inj}^A}$) for 2D,  the dissipation rates $\epsilon^\pm$ [Eqs.~(\ref{eq:epsilon_integral})], $\la E_u \ra/\la E_b\ra$ (ratio of total kinetic and magnetic energies), $\la E^+\ra /\la E^- \ra$, hydrodynamic entropy ($S_H$), and total time $T$ in the units of eddy turnover time. For all the runs, $\epsilon^-_\mathrm{inj}=0.1$, and $\eta = \nu$.}
\begin{tabular}{|>{\centering\arraybackslash}p{0.5cm}|>{\centering\arraybackslash}p{1.3cm}|>{\centering\arraybackslash}p{2cm}|>{\centering\arraybackslash}p{1.3cm}|>{\centering\arraybackslash}p{1cm}|>{\centering\arraybackslash}p{2.3cm}|>{\centering\arraybackslash}p{1cm}|>{\centering\arraybackslash}p{1cm}|>{\centering\arraybackslash}p{1.5cm}|>{\centering\arraybackslash}p{1.6cm}|>
{\centering\arraybackslash}p{0.9cm}|>{\centering\arraybackslash}p{0.7cm}|} 
\toprule
\textbf{}  & Grid Size & $\nu,\eta$ & $\mathrm{Re}$ & $\mathrm{\epsilon^+_{inj} }$ & ${\epsilon_\mathrm{inj}^{H_{M}}}$ (3D) or ${\epsilon_\mathrm{inj}^A}$ (2D) & $\mathrm{\epsilon^+}$ & $\mathrm{\epsilon^-}$ & $\langle E_u \rangle/ \langle E_b \rangle$ & $\langle E^+ \rangle/ \langle E^- \rangle$ & $S_H$ & $T$\\
\toprule
$1$  & \multirow{3}{*}{$8192^2$} & $4.55 \times 10^{-5}$ & $96660$ & $0.1$ & $0.002$ & $0.098$ & $0.097$ & $0.75$  & $1.1$ & 5.19 & $20$\\ 
\cline{1-1} \cline{3-12}
$2$  &  & $6.6 \times 10^{-5}$ & $104700$ & $0.3$ & $0.008$ & $0.272$ & $0.096$ & $0.77$ & $4$ &  4.38 & $24$\\ 
\cline{1-1} \cline{3-12}
$3$  &  & $8 \times 10^{-5}$ & $133500$ & $0.5$ & $0.01$ & $0.521$ & $0.115$ & $0.93$ & $10.4$ & 4.01 &  $25$\\ 
\hline
$4$  &  \multirow{3}{*}{$4096^2$} &  $1.15 \times 10^{-4}$ & $40970$ & $0.1$ & $0.002$ & $0.096$ & $0.096$ & $0.77$  & $1$ & 4.71 & $100$\\ 
\cline{1-1} \cline{3-12}
$5$  &   & $1.65 \times 10^{-4}$  & $41880$ & $0.3$ & $0.008$ & $0.289$ & $0.093$ & $0.75$  & $4.8$ & 4.38 & $100$\\ 
\cline{1-1} \cline{3-12}
$6$  &   & $1.99 \times 10^{-4}$  & $53670$ & $0.5$ & $0.01$ & $0.486$ & $0.10$ & $0.95$  & $10$ &3.77 &  $100$\\ 
\hline
$7$ & \multirow{3}{*}{$1024^3$} & $5.6 \times 10^{-4}$ & $6732$ & $0.1$ & $0$ & $0.096$ & $0.098$ & $0.7$ & $1$ & 4.31 & $5$\\ 
\cline{1-1} \cline{3-12}
$8$ &  & $8.016 \times 10^{-4}$ & $7055$ & $0.3$ & $0$ & $0.303$ & $0.10$ & $1.3$  & $5.5$ & 3.82 & $5$\\ 
\cline{1-1} \cline{3-12}
$9$ &  & $9.2 \times 10^{-4}$ & $7463$ & $0.5$ & $0$ & $0.472$ & $0.102$ & $0.79$ & $11$ & 3.65 & $5$\\ 
\cline{1-1} \cline{1-12}
$10$  &  \multirow{3}{*}{ $512^3$} & $1.75 \times 10^{-3}$ & $2154$ & $0.1$ & $0$ & $0.096$ & $0.097$ & $0.7$  & $0.99$ & 3.65 & $100$\\ 
\cline{1-1} \cline{3-12}
$11$  &   & $2.02 \times 10^{-3}$ & $2332$  & $0.3$ & $0$ & $0.297$ & $0.099$ & $1.3$  & $4.6$ & 3.42 & $100$\\ 
\cline{1-1} \cline{3-12}
$12$  &   & $2.34 \times 10^{-3}$ & $2416$ & $0.5$ & $0$ & $0.479$ & $0.096$ & $1.3$  & $11$ & 3.28 & $100$\\ 
\hline
\end{tabular}
\label{tab:simulation_details}
\end{table*}

In the next section, we present the numerical energy spectra and fluxes, as well as various structure functions.

\section{Numerical Results}
\label{sec:numerical_res}

In this section, we present our numerical results. 

\subsection{Turbulent steady states}
\label{sec:steady_state}

\begin{figure}[htbp!]
\renewcommand{\figurename}{\scriptsize FIG.}
\centering
\includegraphics[scale = 0.85]{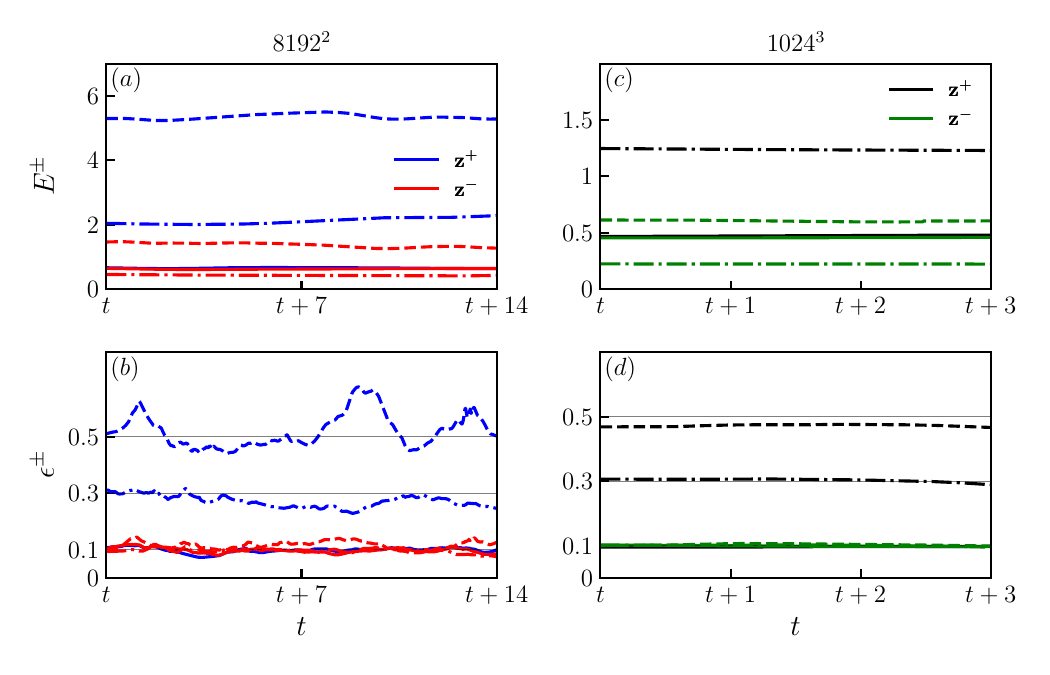}
\caption{\scriptsize For MHD turbulence simulations on $8192^2$ and $1024^3$ grids, time series of  (a,c) $E^\pm$ and (b, d)   $\epsilon^\pm$. For ${\bf z}^+$ and ${\bf z}^-$, blue and red curves are for 2D, respectively, and black and green curves are for 3D, respectively.  Solid, chained, and dashed curves represent $\epsilon_{\mathrm{inj}}^+/\epsilon_{\mathrm{inj}}^-=1$, 3, 5, respectively.}
\label{fig:time_series}
\end{figure}

We performed long simulations, with the total time in the units of eddy turnover time. See Table ~\ref{tab:simulation_details}. In Fig.~\ref{fig:time_series}, we exhibit the time series of global energies $E^\pm$ and the dissipation rates $\epsilon^\pm$ [Eq.~(\ref{eq:epsilon_integral})] for the steady states of $8192^2$ and $1024^3$ runs with $\epsilon^+_{\mathrm{inj}}/\epsilon^-_{\mathrm{inj}}=1,3,5$.  Note the constancy of $E^\pm$ and $\epsilon^\pm$, except for the strong fluctuations in $\epsilon^+$ for the 2D runs, which occur due to the inverse cascade of $A^2$ \cite{Pouquet:JFM1978,Kraichnan:ROPP1980}.  

\begin{figure}[htbp!]
\renewcommand{\figurename}{\scriptsize FIG.}
\centering
\includegraphics[scale = 0.85]{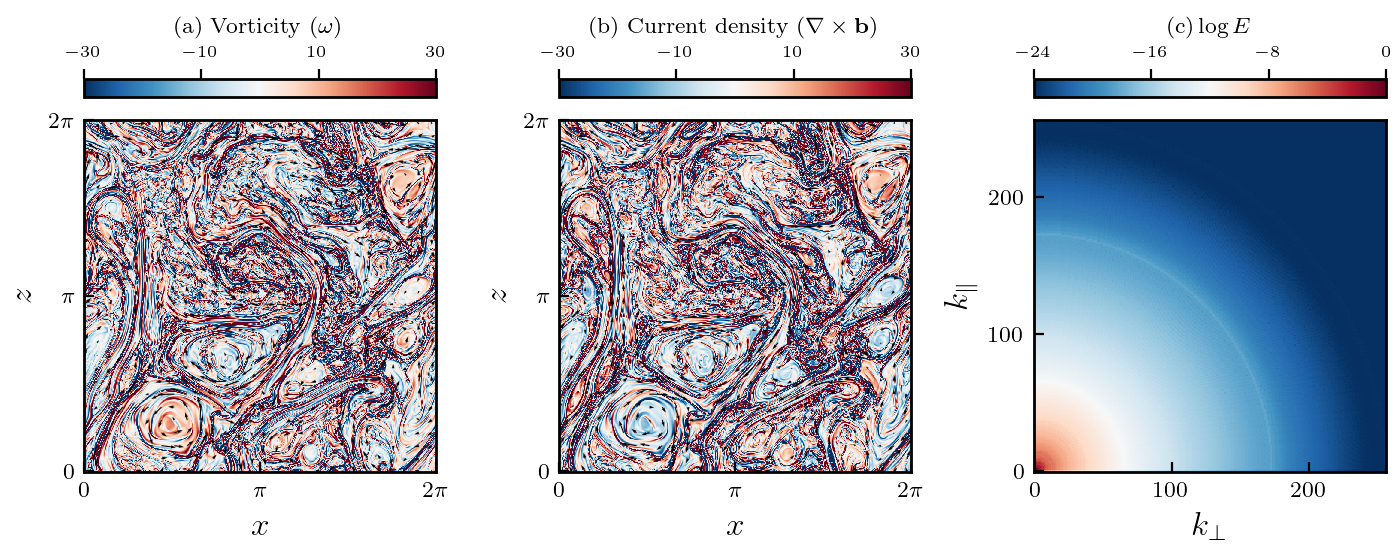}
\caption{\scriptsize For the 2D run on a $8192^2$ grid with $\epsilon_{\mathrm{inj}}^+/\epsilon_{\mathrm{inj}}^-=1$,  (a) density plot of vorticity and vector plot of the velocity field; 
(b) density plot of current density $\nabla \times {\bf b}$ and vector plot of the magnetic field. (c) For the 3D run on a $512^3$ grid with $\epsilon_{\mathrm{inj}}^+/\epsilon_{\mathrm{inj}}^-=1$,  the ring spectrum $E(k,\theta)$ for the total energy.  }
\label{fig:profile_ring}
\end{figure}

In Fig.~\ref{fig:profile_ring}, we exhibit the flow profiles for the $8192^2$ grid simulation and ring spectrum~\cite{Teaca:PRE2009,Sagaut:book,Sundar:PP2017} for the $512^3$ grid simulation, both with $\epsilon^+_{\mathrm{inj}}/\epsilon^-_{\mathrm{inj}}=1$. Figure \ref{fig:profile_ring}(a) illustrates the vorticity as a density plot and the velocity field as a vector plot, whereas Fig.~\ref{fig:profile_ring}(b) illustrates the current density ($\nabla \times {\bf b}$) as a density plot and the magnetic field as a vector plot.  These figures show large-scale structures embedded in a noisy background. Figure~\ref{fig:profile_ring}(c) illustrates the ring spectrum $E(k,\theta)$ for the total energy. The ring spectrum quantifies the energy contents within a ring of radius $k$ at an angle $\theta$, where $\theta$ is the angle between ${\bf B}_0$ and \textbf{k}~\cite{Teaca:PRE2009,Verma:book:ET}. The ring spectrum illustrates that $E(k,\theta) \approx $ constant in $\theta$, indicating the flow to be isotropic.  In Table~\ref{tab:simulation_details}, we list $\epsilon^\pm$, $\langle E_u \rangle/ \langle E_b \rangle$, and $\langle E^+ \rangle/ \langle E^- \rangle$ averaged over the steady state.  The average $\langle E_u \rangle/ \langle E_b \rangle$ ranges from  0.7 to 1.3, whereas $\langle E^+ \rangle/ \langle E^- \rangle$ ranges from 1 to 11. Note that the average dissipation rates $\epsilon^\pm \approx \epsilon_\mathrm{inj}^\pm$, with the maximum relative errors in $\epsilon^\pm $ around $6\%$ in 3D and $10\%$ in 2D.

In the next subsection, we will discuss the energy spectra and fluxes.
 
\begin{table*}
\renewcommand{\tablename}{\scriptsize TABLE}
\caption{\scriptsize Table lists numerical  Kolmogrov's constants $K^\pm$, as well as $ E^+(k)/E^-(k)$, $\Pi^+(k)/\Pi^-(k)$, and the normalized cross helicity $\sigma_c(k)$ averaged over the inertial range.}

\begin{tabular}{|>{\centering\arraybackslash}p{1.5cm}|>{\centering\arraybackslash}p{2cm}|>{\centering\arraybackslash}p{1.6cm}|>{\centering\arraybackslash}p{2.5cm}|>{\centering\arraybackslash}p{2.5cm}|>{\centering\arraybackslash}p{1.5cm}|>{\centering\arraybackslash}p{1.5cm}|>{\centering\arraybackslash}p{2cm}|} 
\toprule
\textbf{}  & Grid Size &  $\mathrm{\epsilon^+_{inj}}$ & $E^+(k)/E^-(k)$ & $\Pi^+(k)/\Pi^-(k)$ & $K^+$ & $K^-$ & $\sigma_c(k)$\\ 
\toprule
$1$  & \multirow{3}{*}{$8192^2$} & $0.1$ & $1.01$ & $0.998$ & $4.4$ & $4.4$ & $0.0031$\\ 
\cline{1-1} \cline{3-8}
$2$  &  & $0.3$ & $4.5$ & $2.7$ & $3.6$ & $5.8$ & $0.61$\\ 
\cline{1-1} \cline{3-8}
$3$  &  & $0.5$ & $4.6$ & $3.1$ & $3.7$ & $7.7$ & $0.65$\\ 
\hline
$4$  &  \multirow{3}{*}{$4096^2$} &  $0.1$  & $1.004$ & $0.995$ & $4.4$ & $4.4$ & $0.0019$\\ 
\cline{1-1} \cline{3-8}
$5$  &   & $0.3$ & $4.2$ & $2.9$ & $3.1$ & $6$ & $0.61$\\ 
\cline{1-1} \cline{3-8}
$6$  &   & $0.5$  & $5.8$ & $4$ & $3$ & $8.2$ & $0.7$\\ 
\hline
$7$ & \multirow{3}{*}{$1024^3$} & $0.1$ & $0.956$ & $0.975$ & $1.8$ & $1.8$ & $-0.02$\\ 
\cline{1-1} \cline{3-8}
$8$ &  & $0.3$ & $4.6$ & $ 2.7$ & $1.4$ & $2.2$ & $0.66$\\ 
\cline{1-1} \cline{3-8}
$9$ &  & $0.5$ & $10.5$ & $4.4$ & $1.5$ & $3.2$ & $0.83$\\ 
\cline{1-1} \cline{1-8}
$10$  &  \multirow{3}{*}{ $512^3$} & $0.1$ & $0.997$ & $1.001$ & $1.8$ & $1.8$ & $-0.0015$\\ 
\cline{1-1} \cline{3-8}
$11$  &   & $0.3$  & $5.2$ & $2.7$ & $1.1$ & $1.7$ & $0.67$\\ 
\cline{1-1} \cline{3-8}
$12$  &   & $0.5$ & $10.3$ & $4.5$ & $1.7$ & $2.4$ & $0.83$\\ 
\hline
\end{tabular}

\label{tab:Kolmogrov_constants}
\end{table*}

\subsection{Energy spectra and fluxes}
\label{sec:energy_spec_flux}

\begin{figure}[htbp!]
\renewcommand{\figurename}{\scriptsize FIG.}
\centering
\includegraphics[scale = 0.85]{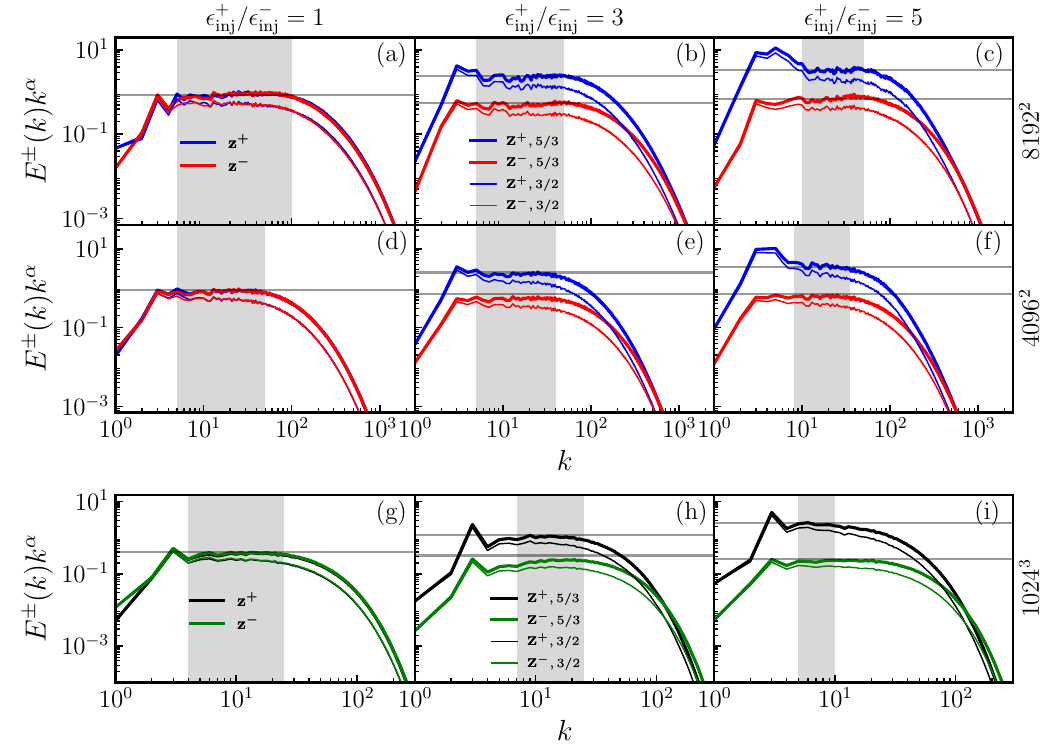}
\caption{\scriptsize Plots of the normalized energy spectra $E^\pm(k) k^\alpha$ with $\alpha =5/3$ and 3/2 on (a,b,c) $8192^2$, (d,e,f) $4096^2$, and (g,h,i) $1024^3$ grids. For 2D, the blue and red curves represent ${\bf z}^+$ and ${\bf z}^-$ fields, respectively. The black and green colors represent the respective curves for 3D.}
\label{fig:spectrum}
\end{figure}

For all the runs, we compute 1D energy spectra and energy fluxes [Eq.~(\ref{eq:flux_def})] by averaging over a number of dataframes. For example, $E^\pm(k)$ and $\Pi^\pm(k)$ for $1024^3$ grids were computed using $3000$ dataframes. In Fig.~\ref{fig:spectrum}, we plot the normalized energy spectra $E^\pm(k)k^{5/3}$ and $E^\pm(k) k^{3/2}$ for $8192^2$, $4096^2$, and $1024^3$ grids, and for $\epsilon^+_\mathrm{inj}/\epsilon^-_\mathrm{inj} = 1$, 3, and 5. As shown in the figure, $E^\pm(k)k^{5/3}$ fit better with the horizontal line than $E^\pm(k)k^{3/2}$. The wavenumber range where $E^\pm(k)k^{5/3} \approx $ const.~(shown as shaded region in Fig.~\ref{fig:spectrum}) is the \textit{inertial range}. In the inertial range, the relative errors in the normalized spectra (relative to the respective constants) are around 10\% to $15\%$ for 2D and around $4\%$ for 3D. We also observe that the inertial range for $\epsilon^+_\mathrm{inj}/\epsilon^-_\mathrm{inj} = 5$ is relatively narrow, which may be due to weak nonlinearity for large cross helicity. Still, the plots in the figure  indicate that $E^\pm(k)$ are in better agreement with Kolmogorov scaling than IK scaling. However, the contrast between the two scalings is not very sharp because the spectral indices $-5/3$ and $-3/2$ are quite close.

\begin{figure}[htbp!]
\renewcommand{\figurename}{\scriptsize FIG.}
\centering
\includegraphics[scale = 0.85]{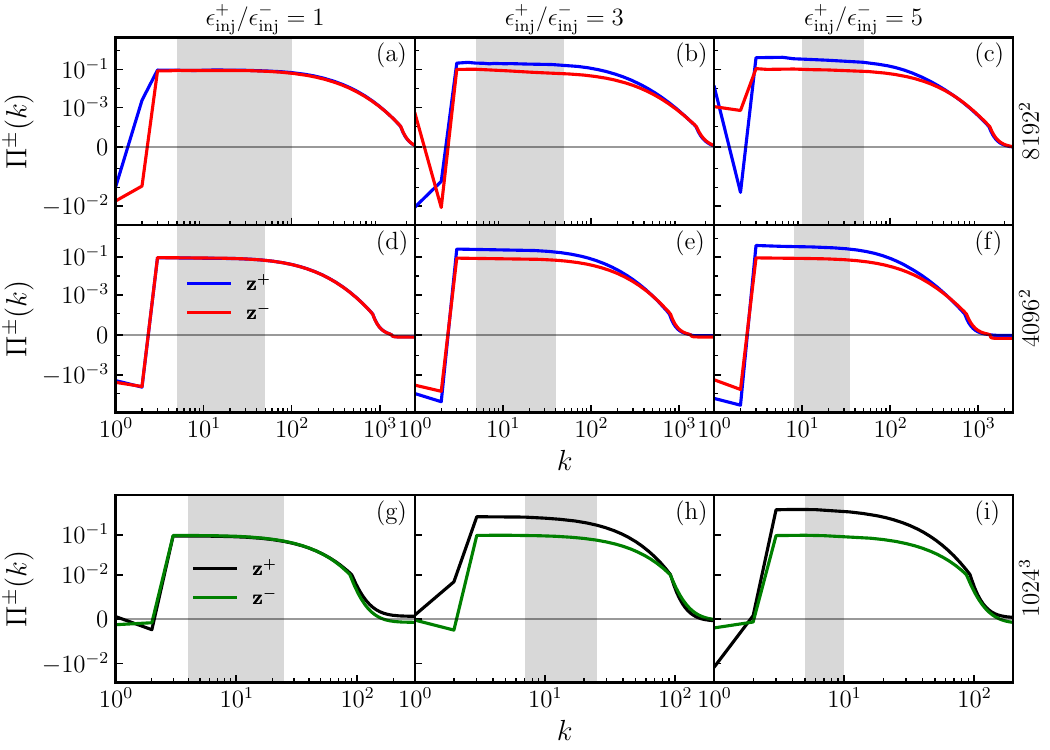}
\caption{\scriptsize Plots of the energy fluxes $\Pi^\pm(k)$ on (a,b,c) $8192^2$, (d,e,f) $4096^2$, and (g,h,i) $1024^3$ grids. For 2D, the blue and red curves represent ${\bf z}^+$ and ${\bf z}^-$ fields, respectively. The black and green colors represent the respective curves for 3D.}
\label{fig:flux}
\end{figure}

In Fig.~\ref{fig:flux}, we plot the energy fluxes $\Pi^\pm(k)$ for $8192^2$, $4096^2$, and $1024^3$ grids. Relative to $E^\pm(k)$, the energy fluxes $\Pi^\pm(k)$ are constant for broader wavenumber bands. More importantly,  for all the \textit{imbalanced} cases, $\Pi^+(k) > \Pi^{-}(k)$  when $E^+(k) > E^-(k)$. These observations are contrary to  Eq.~(\ref{eq:flux_equal}), thus invalidating IK scaling for MHD turbulence. Now, we compare the numerical results with Eq.~(\ref{eq:Kolm_ratio}).  We observe that $E^+(k)/E^-(k) \ne [\Pi^+(k)/\Pi^-(k)]^2$, which necessitates $K^+ \ne K^-$.  Note that $\Pi^+(k)/\Pi^-(k)$ deviates marginally from $\epsilon_\mathrm{inj}^+/\epsilon_\mathrm{inj}^-$. See Table~\ref{tab:Kolmogrov_constants} for details.

\begin{figure}[htbp!]
\renewcommand{\figurename}{\scriptsize FIG.}
\centering
\includegraphics[scale = 0.65]{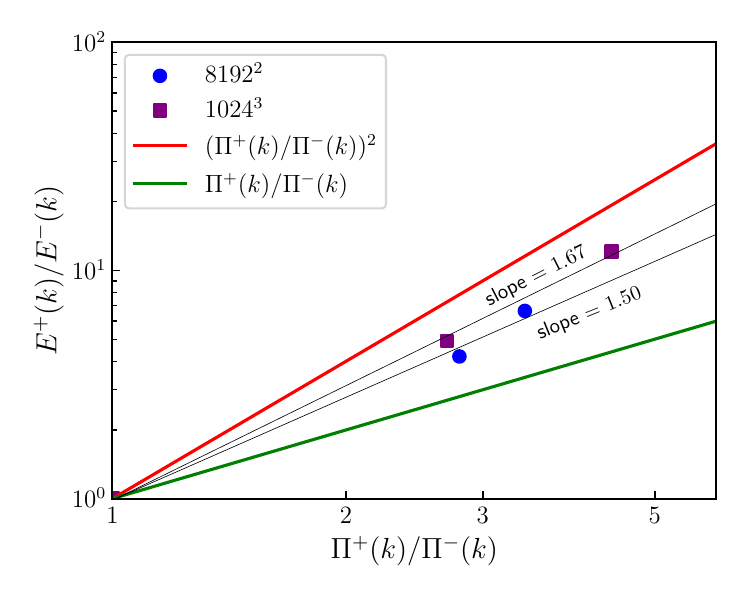}
\caption{\scriptsize Plot of  $E^+(k)/E^-(k)$ vs.~$\Pi^+(k)/\Pi^-(k)$, with $k$ in the inertial range. The blue circles and purple squares are for $8192^2$ and $1024^3$ grids, respectively. The red and green lines represent $(\Pi^+(k)/\Pi^-(k))^2$ and $(\Pi^+(k)/\Pi^-(k))$, respectively. Light black lines represent the best-fit curves with slopes of $1.67$ and $1.5$.}
\label{fig:Ep_by_em_vs_pi}
\end{figure}

We compute Kolmogorov's constant for MHD, $K^\pm$, using the following equations:
\be
K^\pm = \frac{E^\pm(k) k^{5/3}}{(\Pi^\pm)^{4/3}(\Pi^\mp)^{-2/3} }.
\ee
These constants, listed in Table~\ref{tab:Kolmogrov_constants}, range from 3 to 8.2 in 2D, and from 1.2 to 2.8 in 3D. Since $K^- > K^+$ for all the cases, we attempt to fit  \be
\frac{E^+(k)}{E^-(k)}  = \left[  \frac{\Pi^+(k)}{\Pi^-(k)}  \right]^\beta
\ee
with $\beta <2$. In Fig.~\ref{fig:Ep_by_em_vs_pi}, we plot $E^+(k)/E^-(k)$ vs.~$\Pi^+(k)/\Pi^-(k)$ for the imbalanced 2D and 3D runs. 
Using the best-fit curves, we deduce that
\be 
\frac{E^+(k)}{E^-(k)}  \approx \left\{ 
\begin{aligned}
&\left[  \frac{\Pi^+(k)}{\Pi^-(k)}  \right]^{1.5}  \text{for $d=2$} \\
&\left[  \frac{\Pi^+(k)}{\Pi^-(k)}  \right]^{1.67}  \text{for $d=3$}.
\end{aligned}
\right.
\label{eq:Ek_Pik_reln}
\ee
We remark that the relation of Eq.~(\ref{eq:Ek_Pik_reln}) may not hold when $\sigma_c \rightarrow 1$, which may be a different turbulence regime with weak nonlinearity. \\

The above observations on $\Pi^\pm(k)$ support Kolmogorov scaling quite strongly. Earlier, \citet{Verma:JGR1996DNS} had adopted a similar strategy, but their grid resolutions, $512^2$ for 2D and $128^3$ for 3D, were too coarse. The present study with sufficient grid resolutions is conclusive.

\subsection{Structure functions and intermittency exponents}
\label{sec:Sp}

In this subsection, we present numerical results on the third-order structure functions and  intermittency exponents and compare them with the predictions of IK and Kolmogorov-like models. Refer to Sec.~\ref{sec:Theory} for the predictions on the structure functions by IK and Kolmogorov-like models.

\begin{figure}[htbp!]
\renewcommand{\figurename}{\scriptsize FIG.}
\centering
\includegraphics[scale = 0.85]{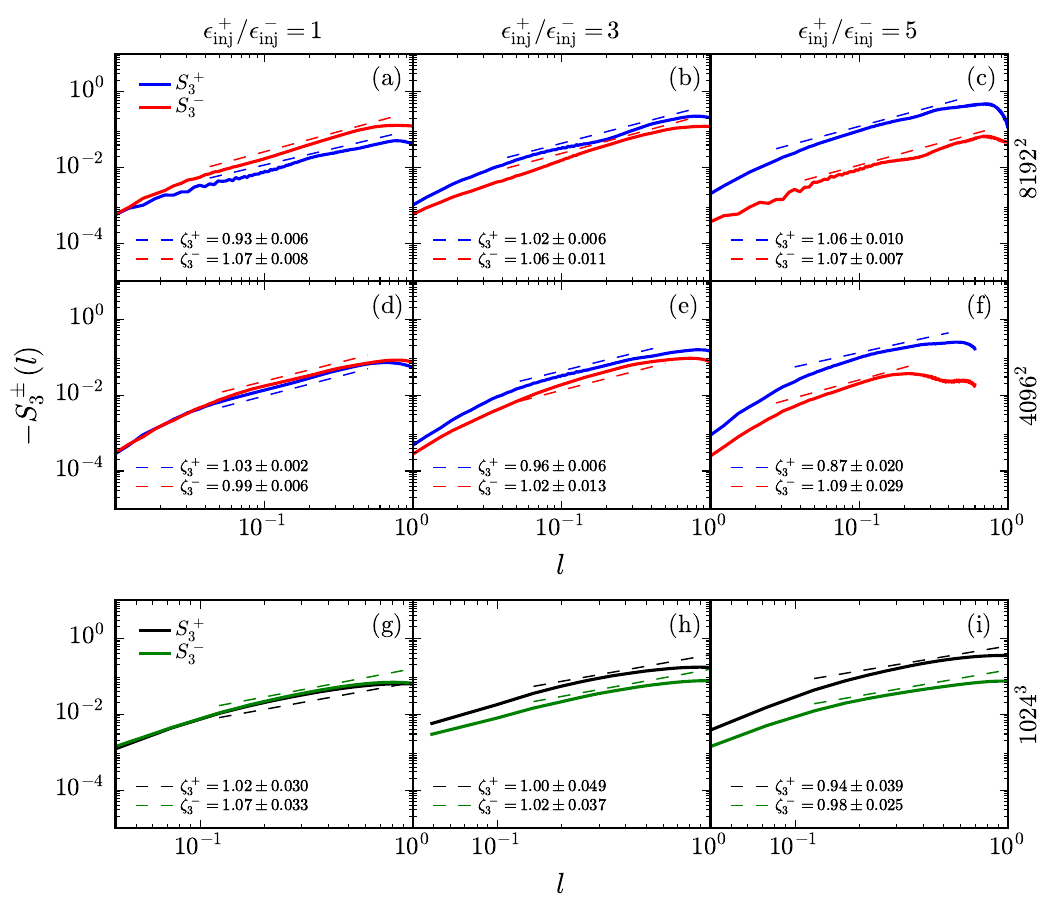}
\caption{\scriptsize Plots of the time averaged third-order structure functions $-S_3^\pm(l)$ for  2D runs on $8192^2$ grid (a,b,c) and on $4096^2$ grid (d,e,f); and for 3D runs on $1024^3$ grid (g,h,i).  Red and green curves are for 2D, whereas the black and green curves are for 3D.  The dashed lines represent the best-fit curves.}
\label{fig:SF3_combo}
\end{figure}

Using the numerical data, we compute $S^\pm_3(l)$ defined in Eq.~(\ref{eq:S3}). To suppress fluctuations, we average $S^\pm_3(l)$ over 5 dataframes in 2D and 3 dataframes in 3D. In Fig.~\ref{fig:SF3_combo} we plot the   $-S^\pm_3(l)$ for $8192^2$, $4096^2$, and $1024^3$ grids and for $\epsilon^+_\mathrm{inj}/\epsilon^-_\mathrm{inj} = 1$, 3, and 5. We observe that
\be
S^\pm_3(l) \propto l^{\zeta_3^\pm}.
\label{eq:zeta3}
\ee
We compute $\zeta_3^\pm$ using the best-fit curves to $S_3^\pm(l)$ vs.~$l$ plots and list them in Table~\ref{tab:zeta3} for the $8192^2$ and $1024^3$ grids. We observe that $\zeta_3^\pm \approx 1$, thus validating Kolmogorov scaling for MHD turbulence.  We also observe that 
\be
\frac{S^\pm_3(l)  d}{4l}  \approx -\epsilon^\pm
\label{eq:S3_eps_reln}
\ee
for all the runs, which is consistent with Eq.~(\ref{eq:S3}). Note, however, that Eq.~(\ref{eq:S3_eps_reln}) deviates somewhat for 2D due to the strong fluctuations. 

\begin{table*}
\renewcommand{\tablename}{\scriptsize TABLE}
\caption{\parbox[t]{\linewidth}{\raggedright\scriptsize Numerically computed $\zeta_3^\pm$ [Eq.~(\ref{eq:zeta3})] for 2D runs on $8192^2$ grid and for 3D runs on $1024^3$ grid. }}
\renewcommand{\arraystretch}{1.2}
\setlength{\tabcolsep}{8pt} 
\resizebox{\textwidth}{!}{ 
\begin{tabular}{|c|c|c|c|c|c|c|c|c|c|}
        \hline
        $\epsilon_{\mathrm{inj}}^+ / \epsilon_{\mathrm{inj}}^-$& 1 (2D) & 3 (2D) &  5 (2D)  & 1 (3D) & 3 (3D) & 5 (3D) \\  
        \hline 
        $\zeta_3^{+}$ & $0.93 \pm 0.006$ & $1.02 \pm 0.006$ & $1.06 \pm 0.01$ & $1.02 \pm 0.030$ & $1.00 \pm 0.049$ & $0.94 \pm 0.039$ \\  
        \hline
        $\zeta_3^{-}$ & $1.07 \pm 0.008$ & $1.06 \pm 0.011$ & $1.07 \pm 0.007$ & $1.07 \pm 0.033$ & $1.02 \pm 0.037$ & $0.98 \pm 0.025$ \\  
        \hline
    \end{tabular}
    }
    \label{tab:zeta3}
\end{table*}

\begin{figure}[htbp!]
\renewcommand{\figurename}{\scriptsize FIG.}
\centering
\includegraphics[scale = 0.85]{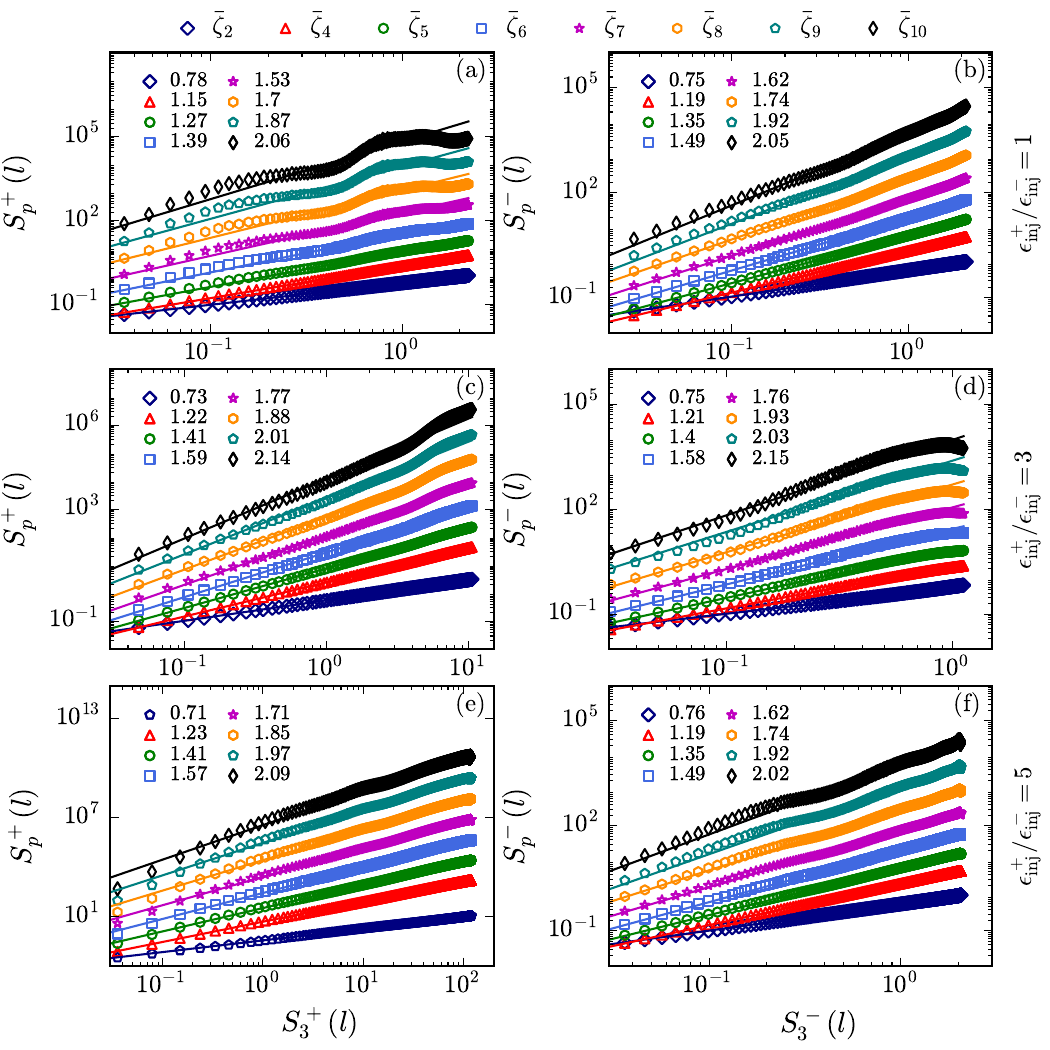}
\caption{\scriptsize For 2D runs with $8192^2$ grids, the plots of averaged structure functions $S_p^\pm(l)$ vs.~$S_3^\pm(l)$ for (a,b) $\epsilon_{\mathrm{inj}}^+/\epsilon_{\mathrm{inj}}^- = 1$, (c,d) $\epsilon_{\mathrm{inj}}^+/\epsilon_{\mathrm{inj}}^- = 3$, and (e,f) $\epsilon_{\mathrm{inj}}^+/\epsilon_{\mathrm{inj}}^- = 5$. Left and right columns are for $\mathbf{z^+}$ and $\mathbf{z^-}$ fields, respectively.  }
\label{fig:ess_2D}
\end{figure}

\begin{figure}[htbp!]
\renewcommand{\figurename}{\scriptsize FIG.}
\centering
\includegraphics[scale = 0.85]{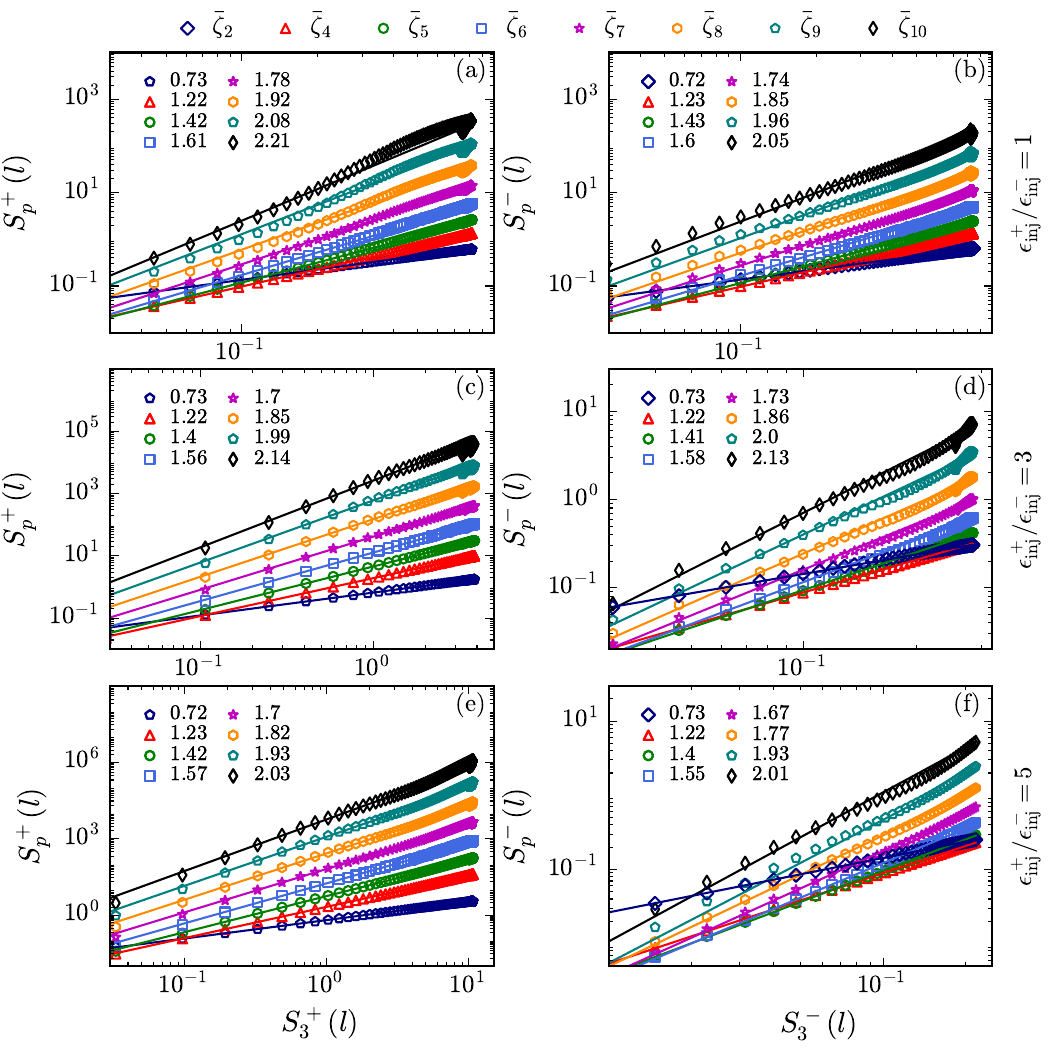}
\caption{\scriptsize For 3D runs with $1024^3$ grids, the plots of the averaged structure functions $S_p^\pm(l)$ vs.~$S_3^\pm(l)$ for (a,b) $\epsilon_{\mathrm{inj}}^+/\epsilon_{\mathrm{inj}}^- = 1$, (c,d) $\epsilon_{\mathrm{inj}}^+/\epsilon_{\mathrm{inj}}^- = 3$, and (e,f) $\epsilon_{\mathrm{inj}}^+/\epsilon_{\mathrm{inj}}^- = 5$. Left and right columns are for $\mathbf{z^+}$ and $\mathbf{z^-}$ fields, respectively. }
\label{fig:ess_3D}
\end{figure}

\begin{table}[h!]
    \renewcommand{\tablename}{\scriptsize TABLE}
    \caption{\parbox[t]{\linewidth}{\raggedright\scriptsize The intermittency exponents $\bar{\zeta}_p^\pm$  for Runs 1 to 9 listed in  table~\ref{tab:simulation_details}. The table covers $8192^2, 4096^2$, and $1024^3$ grids.}}
    \centering
    \begin{tabular}{|>{\centering\arraybackslash}p{0.8cm}|>{\centering\arraybackslash}p{0.8cm}|>{\centering\arraybackslash}p{0.8cm}|>{\centering\arraybackslash}p{0.8cm}|>{\centering\arraybackslash}p{0.8cm}|>{\centering\arraybackslash}p{0.8cm}|>{\centering\arraybackslash}p{0.8cm}|>{\centering\arraybackslash}p{0.8cm}|>{\centering\arraybackslash}p{0.8cm}|>{\centering\arraybackslash}p{0.8cm}|>{\centering\arraybackslash}p{0.8cm}|>{\centering\arraybackslash}p{0.8cm}|>{\centering\arraybackslash}p{0.8cm}|>{\centering\arraybackslash}p{0.8cm}|>{\centering\arraybackslash}p{0.8cm}|>{\centering\arraybackslash}p{0.8cm}|>{\centering\arraybackslash}p{0.8cm}|}
    \toprule
     Run& \multicolumn{2}{c|}{$p=2$}& \multicolumn{2}{c|}{$p=4$} & \multicolumn{2}{c|}{$p=5$}& \multicolumn{2}{c|}{$p=6$} & \multicolumn{2}{c|}{$p=7$} & \multicolumn{2}{c|}{$p=8$} & \multicolumn{2}{c|}{$p=9$} & \multicolumn{2}{c|}{$p=10$} \\ \hline
    \hline
       & $\bar{\zeta}_p^+$ & $\bar{\zeta}_p^-$  & $\bar{\zeta}_p^+$ & $\bar{\zeta}_p^-$ & $\bar{\zeta}_p^+$ & $\bar{\zeta}_p^-$ & $\bar{\zeta}_p^+$ & $\bar{\zeta}_p^-$ & $\bar{\zeta}_p^+$ & $\bar{\zeta}_p^-$ & $\bar{\zeta}_p^+$ & $\bar{\zeta}_p^-$ & $\bar{\zeta}_p^+$ & $\bar{\zeta}_p^-$ & $\bar{\zeta}_p^+$\rule{0pt}{2.5ex} & $\bar{\zeta}_p^-$  \\ \hline
    1  &0.78& 0.75 & 1.15& 1.19 & 1.28& 1.35 & 1.39& 1.49 & 1.53& 1.62 & 1.7& 1.74 & 1.87& 1.92 & 2.06& 2.05  \\ \hline
    2  & 0.73 & 0.75 &1.22& 1.21  &1.41& 1.40  &1.59& 1.58  &1.77& 1.76  &1.88& 1.93  &2.01& 2.03  & 2.14& 2.15  \\ \hline
    3  &0.71& 0.76  & 1.23& 1.19 &1.41& 1.35  &1.57& 1.49  &1.71& 1.62  &1.85& 1.74  &1.97& 1.92  &2.09& 2.02   \\ \hline
    
    4  &0.74& 0.74 &1.21& 1.21 &1.38& 1.38 &1.54& 1.53  &1.68& 1.67 &1.81& 1.8  &1.93& 1.93  & 2.05& 2.06  \\ \hline
    5  &0.75& 0.76 &1.2& 1.18 &1.37& 1.34 &1.52& 1.47 &1.66& 1.59 &1.8& 1.72 &1.93& 1.84 & 2.07& 1.97  \\ \hline
    6  &0.72& 0.73 &1.22& 1.21 &1.41& 1.38 &1.58& 1.54 &1.73& 1.68 &1.88& 1.82 &2.03& 1.96  & 2.19& 2.09  \\ \hline
    
    7 &0.73& 0.72  &1.22& 1.23 &1.42& 1.43  &1.61& 1.60  &1.78& 1.74  &1.92& 1.85  &2.08& 1.96  &2.21& 2.05   \\ \hline
    8  &0.73& 0.73  &1.22& 1.22  &1.40& 1.41  &1.56& 1.58  &1.70& 1.73  &1.85& 1.86 &1.99& 2.00  & 2.14& 2.13  \\ \hline
    9  &0.72& 0.73  &1.23& 1.22  &1.42& 1.40  &1.57& 1.55  &1.70& 1.67  &1.82& 1.77  &1.93& 1.93  &2.03& 2.01   \\ \hline
    
    \end{tabular}
    \label{tab:ess}
\end{table}

Next, we numerically compute the structure functions for various orders [Eq.~(\ref{eq:Sp})]. We average over 8 dataframes in 2D and 3 in 3D. We can compute the intermittency exponent $\zeta_p^\pm$ using best-fit curve to the $S_p^\pm(l)$ vs.~$l$ plots. However,  \textit{Extended Self-Similarity} (ESS) \cite{Benzi:PRE1993} broadens the scaling range and improves the accuracy of $\zeta_p^\pm$. Since $S^\pm_3(l) \propto l$  for MHD, we determine $\zeta_p^\pm$ using the best-fit cuvers to the $S_p^\pm(l)$ vs.~$S_3^\pm(l)$  plots, which  yields
\be
\bar{\zeta}^\pm_{p} = \frac{\zeta^\pm_p}{\zeta^\pm_3} \approx \zeta_p^\pm.
\ee

In Figs.~\ref{fig:ess_2D} and \ref{fig:ess_3D}, we plot $S^\pm_p(l)$ vs.~$S^\pm_3(l)$ for $8192^2$  and $1024^3$ grids, respectively. The three rows in the figure represent  $\epsilon^+_\mathrm{inj}/\epsilon^-_\mathrm{inj} = 1$, 3, and 5. Linear fits in log-log plots confirm the validity of ESS.  The intermittency exponents, $\bar{\zeta}^\pm_{p}$, for Runs 1 to 9 of Table~\ref{tab:simulation_details} (grids $8192^2, 4096^2$, and $1024^3$) are listed in Table \ref{tab:ess}.  The errors in $\bar{\zeta}^\pm_{p}$ are $O(0.01)$.  For example. the maximum error in $\tilde{\zeta}_p^\pm$ for the $8192^2$ grid is $0.02$ and that for the $1024^3$ grid is $0.03$. Note that $\tilde{\zeta}_4^\pm$ is far from 1, which is contrary to the predictions of IK scaling [Eq.~(\ref{eq:Sp_IK})].

\begin{figure}[htbp!]
\renewcommand{\figurename}{\scriptsize FIG.}
\centering
\includegraphics[scale = 0.85]{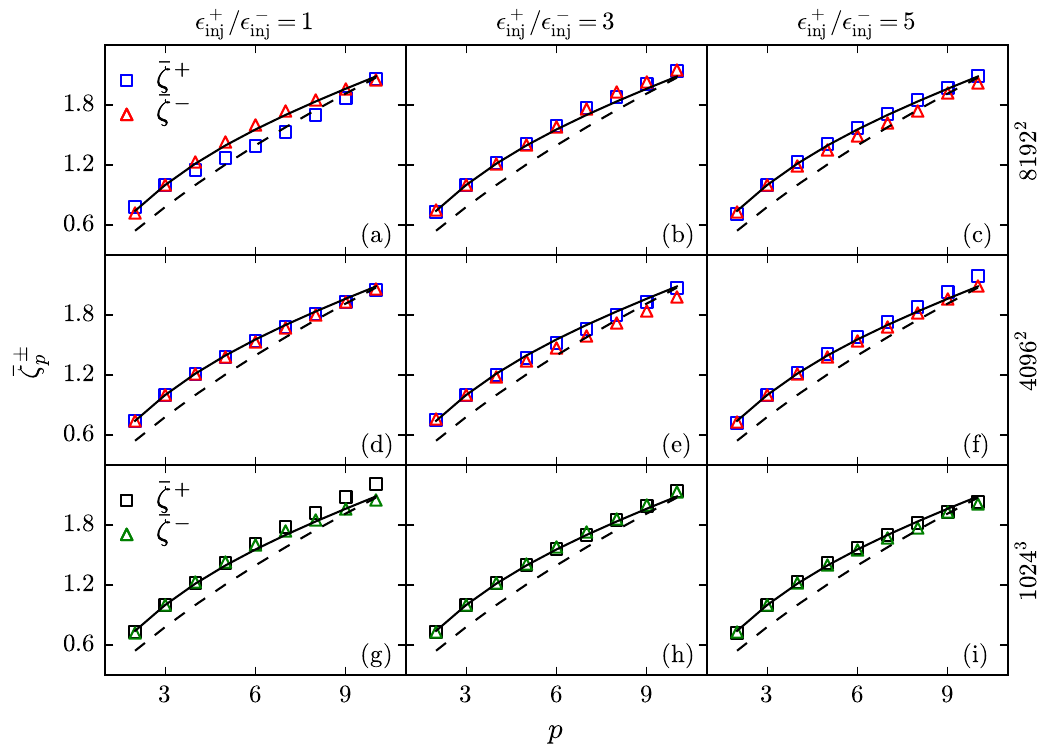}
\caption{\scriptsize Plots of the intermittency exponents, $\bar{\zeta}_p^\pm$. (a,b,c): 2D runs on $8192^2$ grids, (d,e,f): 2D runs on  $4096^2$ grids, and (g,h,i): 3D runs on $1024^3$ grids. The blue and black squares represent $\bar{\zeta}_p^+$ for 2D and 3D, respectively; whereas the red and green triangles represent the respective $\bar{\zeta}_p^-$'s. The solid and dotted curves depict the Kolmogorov and IK scaling, respectively.}
\label{fig:zeta_combo}
\end{figure}

In Fig.~\ref{fig:zeta_combo}, we plot $\bar{\zeta}^\pm_{p}$ vs.~$p$ for $8192^2$, $4096^2$, and $1024^3$ grids. In the figure, the squares represent $\bar{\zeta}^+_{p}$, whereas the triangles represent $\bar{\zeta}^-_{p}$. The solid  and dashed curves represent the Kolmogorov scaling [Eq.~(\ref{eq:Sp_Kolm})] and IK scaling [Eq.~(\ref{eq:Sp_IK})], respectively. We observe that the numerical results are closer to the solid curves than the dashed curves, indicating strong support for the Kolmogorov scaling. The 2D $\bar{\zeta}^+_{p}$ curves deviate marginally from Eq.~(\ref{eq:Sp_Kolm}), which may be related to the strong fluctuations in 2D turbulence.


In summary,  our numerical results on the energy spectra and fluxes, third-order structure functions, and intermittency exponents support Kolmogorov scaling.

\section{Comparison with past results on numerical simulations and solar wind observations}
\label{sec:past}

In early attempts,  Biskamp, Welter, M\"{u}ller, and Schwarz~\cite{Biskamp:PFB1989,Muller:PRL2000,Biskamp:PP2001,Biskamp:PP2000} simulated 2D and 3D MHD turbulence on $4096^2$ and $512^3$ grids and showed that for 3D MHD turbulence,  $E_u(k), E_b(k) \sim  k^{-5/3}$, and  the multiscaling exponents $\zeta_p$'s follow Eq.~(\ref{eq:Sp_Kolm}). \citet{Sorriso:PP2002} and \citet{Mininni:PRE2009a} too reported $\zeta_3 \approx 1$. These results are consistent with those presented in this paper.  For 2D MHD turbulence, Biskamp, Welter, M\"{u}ller, and Schwarz~\cite{Biskamp:PFB1989,Biskamp:PP2000} argued that $E_u(k)$ and $E_b(k)$ follow IK scaling, but $\zeta_3 \approx 1$. In addition, $\zeta_p$ (other than $p=3$) are inconclusive on 2D MHD turbulence.  Our numerical results on 2D MHD turbulence support Kolmogorov spectrum, thus counter the past results of Biskamp and coworkers.    \citet{Verma:JGR1996DNS} simulated  MHD turbulence on coarse grids ($512^2$ and $128^3$) and observed agreement with Eq.~(\ref{eq:Kolm_ratio}), thus validating Kolmogorov scaling for MHD turbulence (also, see \cite{Verma:PP2002}).  The numerical results of \citet{Sahoo:NJP2011} lead to mixed conclusions.

Since 2006, most numerical works have been for anisotropic turbulence.   \citet{Cho:ApJ2000} and \citet{Beresnyak:PRL2011,Beresnyak:LR2019} verified Goldreich-Sridhar  model~\cite{Goldreich:ApJ1995}, whereas \citet{Mason:PRL2006,Mason:PRE2008} and \citet{Podesta:PP2010}  found agreement with \citet{Boldyrev:PRL2006}'s predictions.  Also see \citet{Sundar:PP2017} for the energy flux studies in anisotropic MHD.   Often, these researchers simulated reduced magnetohydrodynamics (RMHD) with ${\bf B}_0$.  In this section, we compare our isotropic results with those for anisotropic ones, even though a comparison is not fully justified.

\begin{figure}[htbp!]
\renewcommand{\figurename}{\scriptsize FIG.}
\centering
\includegraphics[scale = 0.85]{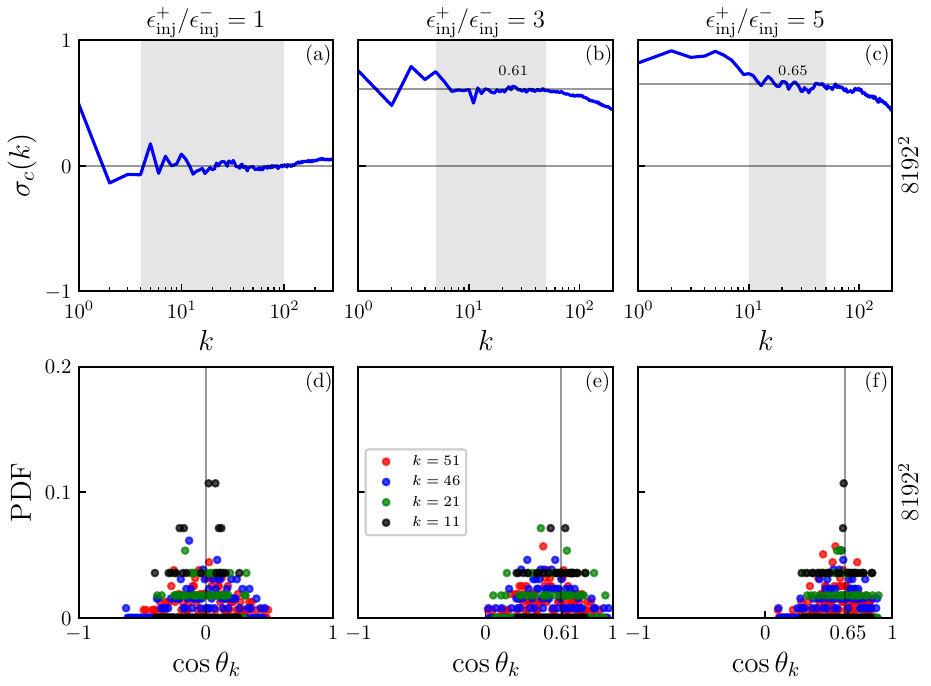}
\caption{\scriptsize For the 2D runs on $8192^2$ grid with $\epsilon^+_{\mathrm{inj}}/\epsilon^-_{\mathrm{inj}} = 1, 3,$ and $5$,  (a,b,c) plots of $\sigma_c(k)$, and (d,e,f) plots of probability density function (PDF)  of  $\cos\theta_k$ for $k=11, 21, 46$, and 51.}
\label{fig:cos_theta_k}
\end{figure}

\begin{figure}[htbp!]
\renewcommand{\figurename}{\scriptsize FIG.}
\centering
\includegraphics[scale = 0.85]{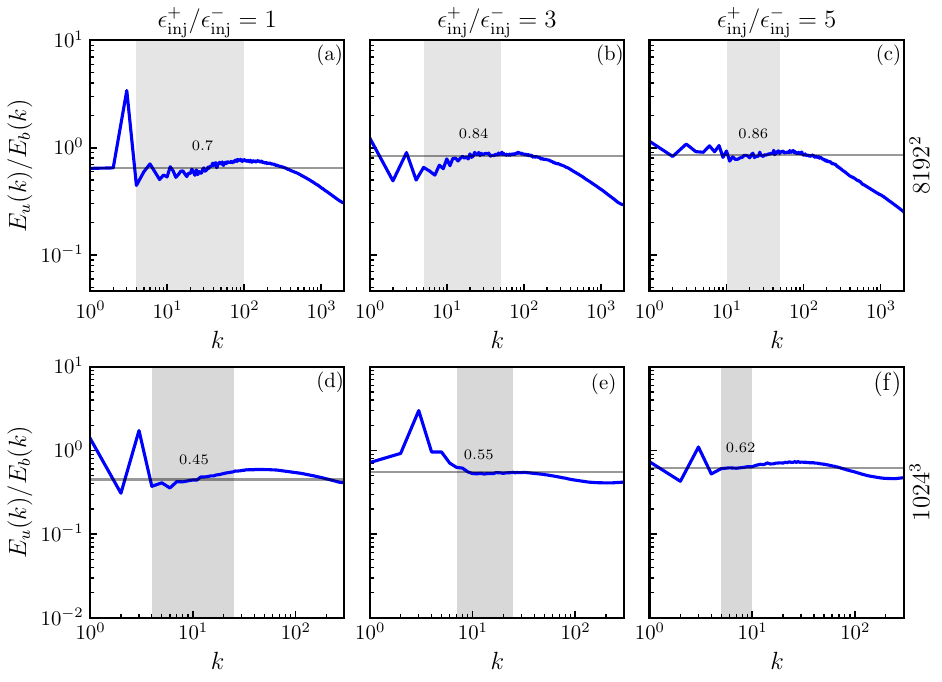}
\caption{\scriptsize Plots of the Alfv\'en ratio ($E_u(k)/E_b(k)$) with $\epsilon^+_{\mathrm{inj}}/\epsilon^-_{\mathrm{inj}} = 1, 3,$ and $5$, (a,b,c) plots for $8192^2$, and (d,e,f) plots for $1024^3$.}
\label{fig:Alfven_ratio}
\end{figure}

Researchers \cite{Matthaeus:JGR1982rugged,Tu:SSR1995,Podesta:PP2010,Parashar:PRL2018} have computed the normalized cross helicity spectrum, $\sigma_c(k) = [E^+(k)-E^-(k)]/[E^+(k)+E^-(k)] $, for the solar wind data and showed it to be nearly constant in the inertial range, which is consistent with Eq.~(\ref{eq:Kolm}). Note that $\sigma_c(k)= \cos \theta_k$, both of which are averaged over the wavenumber shell of radius $k$. In Figure~\ref{fig:cos_theta_k}(a,b,c) we plot $\sigma_c(k)$ for the $8192^2$ grid simulations with $\epsilon_{\mathrm{inj}}^+/\epsilon_{\mathrm{inj}}^- = 1, 3$, and 5, respectively. In the three plots, $\sigma_c(k) \approx 0, 0.6, 0.65$ in the inertial range, whereas $\sigma_c(k)$ is larger than the mean for smaller $k$'s but less than the mean for larger $k$'s; these results are consistent with the reported $\sigma_c(k)$ for the solar wind~\cite{Matthaeus:JGR1982rugged,Tu:SSR1995,Podesta:PP2010,Parashar:PRL2018}. We complement the above plots with the probability density functions (PDF)  of  $\cos\theta_k$ for $k=11, 21, 46$, and 51. The plots show that the PDFs peak at the respective mean values, with significant fluctuations around the mean. Consistency and nonzero values of $\sigma_c(k)$ in the inertial range is consistent with Kolmogorov scaling [Eq.~(\ref{eq:Kolm})]. Note that $\sigma_c(k) \approx 0$ in IK scaling because of $\Pi^+(k) \approx \Pi^-(k)$.
We remark that the above $\sigma_c(k)$ or $\theta_k$ should not be compared with the  predictions of \citet{Boldyrev:PRL2006}, which are for anisotropic MHD turbulence. For completeness, in Fig.~\ref{fig:Alfven_ratio}, we also plot the Alfv\'{e}n ratio $r_A(k) = E_u(k)/E_b(k)$  for $8192^2$ and $1024^3$ grids. For $\epsilon^+_{\mathrm{inj}}/\epsilon^-_{\mathrm{inj}} = 1, 3,$ and $5$, the average values of $r_A(k)$ for $8192^2$ grid are 0.7, 0.84, and 0.86, respectively, whereas those for the $1024^3$ grid are 0.45, 0.55, and 0.62, respectively. 

\citet{Marsch:RMA1991} and \citet{Lithwick:ApJ2007}  predicted that $E^+(k)/E^-(k) \approx [\Pi^+/\Pi^-]^2 $, whereas \citet{Beresnyak:ApJ2008a} and \citet{Perez:PRL2009} predicted that $E^+(k)/E^-(k) \approx \Pi^+/\Pi^- $.  However, we observe scaling in between: $E^+(k)/E^-(k) \approx [\Pi^+/\Pi^-]^{1.5} $ for 2D and $E^+(k)/E^-(k) \approx [\Pi^+/\Pi^-]^{1.67} $ for 3D; the divergence from  $E^+(k)/E^-(k) \approx [\Pi^+/\Pi^-]^2 $ is attributed to  $K^+ \ne K^-$.  Also, the Kolmogorov's constants reported by us are in general agreement (within a factor of 2) with the past simulations, e.g., \citet{Beresnyak:PRL2011,Beresnyak:LR2019} and field-theoretic computations~\cite{Verma:PR2004,Verma:arxiv2025FT}.

Solar wind too provides a platform to test the validity of MHD turbulence models.  As described in the introduction (Sec.~\ref{sec:intro}), some solar wind observations, especially earlier ones, support Kolmogorov's $k^{-5/3}$ spectrum~\cite{Taylor:PRS1938,Matthaeus:JGR1982rugged,Tu:SSR1995}. However, some later works provide support to IK scaling. For example, \citet{Podesta:PP2010} reported   $k^{-3/2}$ spectrum for large $\sigma_c$, but   $k^{-5/3}$ spectrum when $\sigma_c \approx 0$. In addition, \citet{Podesta:ApJ2007} reported $k^{-5/3}$ spectrum for the magnetic field and $k^{-3/2}$ spectrum for the velocity field. In the companion Letter~\cite{Verma:MHD_PRL}, we attribute this divergence to the energy transfers from \textbf{b} to \textbf{u} in the inertial range~\cite{Verma:Fluid2021}. Also, solar wind observations~\cite{Tu:SSR1995,Podesta:PP2010,Parashar:PRL2018} tend to support $\cos \theta_k = \sigma_c(k) \approx $ const., contradictory to \citet{Boldyrev:PRL2006}'s predictions. Regarding the structure functions, \citet{SorrisoValvo:PRL2007} reported that $S_3(l) \propto l$, which is consistent with Eq.~(\ref{eq:S3}).   However, \citet{Bruno:LR2013}'s intermittency exponents are inconsistent with both  Kolmogorov and IK scalings [Eqs.~(\ref{eq:Sp_IK}, \ref{eq:Sp_Kolm})].

For strong turbulence, several authors, including \citet{Kraichnan:PF1965MHD}, argued that the large-scale magnetic field would play the role of mean magnetic field, which would lead to  IK scaling [Eq.~(\ref{eq:IK})]. This is contrary to the results presented in this paper and in some previous works.  \citet{Verma:PP1999} attempted to reconcile this ambiguity using renormalization group analysis of mean magnetic field.  He showed that the \textit{renormalized mean magnetic field} $B_0(k)$ scales as $(\epsilon^T)^{1/3} k^{-1/3}$, substitution of which in Eq.~(\ref{eq:IK}) leads to Kolmogorov spectrum, Eq.~(\ref{eq:Kolm}),  for strong MHD turbulence~\cite{Verma:book:ET}. These arguments need to be tested numerically. Also note that  renormalization group analysis  and energy flux calculations mostly support Kolmogorov scaling \cite{Verma:PP1999,Verma:PRE2001,Verma:PR2004,Adzhemyan:book:RG,Zhou:PR2010,Verma:arxiv2025FT}, but there are variants.

We conclude in the next section.

\section{Conclusions}
\label{sec:Conclusions}
Accurate modelling of MHD turbulence has remained uncertain even after 60 years of research.  This paper addresses this problem and numerically shows that Kolmogorov scaling is valid for \textit{isotropic} MHD turbulence.  Since 2D and 3D MHD turbulence have similar turbulence dynamics, we perform high resolution simulations ($8192^2$ and $4096^2$) in 2D and moderately-resolved simulations ($1024^3$ and $512^3$) in 3D.  Since the inertial-range energy fluxes  of the Els\"{a}sser variables, $\Pi^\pm(k)$, are constant, we focus on these variables, rather than velocity and magnetic fields, which have variable energy fluxes in the inertial range~\cite{Verma:MHD_PRL}. We limit ourselves to isotropic turbulence with ${\bf B}_0 = 0$. Hence, our results may differ in detail from those related to the  anisotropic models proposed by \citet{Goldreich:ApJ1995} and \citet{Boldyrev:PRL2006}.

The main results of our numerical simulations are as follows:
\begin{enumerate}
    \item Our numerically-computed $E^\pm(k)$ are closer to $k^{-5/3}$ than $k^{-3/2}$. However, the spectral indices do not provide conclusive evidence because they are too close.

    \item The IK and Kolmogorov frameworks have different predictions for the inertial-range energy fluxes $\Pi^\pm(k)$. Using these diagnostics, we show that numerical energy fluxes support Kolmogorov framework over IK framework.

    \item Isotropic MHD turbulence has an exact relationship: $S^\pm _3(l) = -(4/3) \epsilon^\pm l$,  which is consistent with the Kolmogorov scaling. In contrast, IK scaling predicts that $S^\pm _4(l) \propto l$. 
    Our numerical simulation clearly support Kolmogorov scaling.

    \item Numerically-computed intermittency exponents support Kolmogorov scaling.
    
\end{enumerate} 
In the companion Letter~\cite{Verma:MHD_PRL}, we show that the total energy and cross helicity too exhibit spectra, fluxes, and third-order structure functions, consistent with Kolmogorov scaling.  In contrast, $E_u(k) \sim k^{-3/2}$ and $E_b(k) \sim k^{-5/3}$ due to energy transfer from \textbf{b} to \textbf{u}. 

Thus, our novel approach resolves a long-standing problem of MHD turbulence. The present paper provides many valuable results and insights into isotropic MHD turbulence. We believe similar investigations of \textit{anisotropic MHD turbulence} using high-resolution simulations would be very useful. A clear understanding of isotropic and anisotropic MHD turbulence would be a major step in turbulence research, and it will open a window for a better modeling of solar-, geo-, and galactic dynamos, solar wind, solar corona, etc.

\begin{acknowledgments}
We thank Melvyn Goldstein, Aaron Roberts, William Matthaeus, Riddhi Bandyopadhyay, Jayant Bhattacharjee, Alexander Schekochihin, Stephan Fauve, Alex Alexakis, Annick Pouquet, Rodion Stepanov, Franck Plunian, and Soumyadeep Chatterjee for valuable suggestions at various stages of our investigation. This research used resources of the Oak Ridge Leadership Computing Facility (through Director’s Discretionary Program) at the Oak Ridge National Laboratory.    Simulations were also performed on Param Sanganak (IIT Kanpur), and HPC facility of Kotak School of Sustainability (KSS). Part of this work
was done in the Center for Turbulence Research, Stanford University, where MKV was a Visiting Senior Fellow.  Part of this work was supported
by Anusandhan National Research Foundation, India (Grant Nos.
SERB/PHY/2021522 and SERB/PHY/2021473), and the J. C. Bose
Fellowship (SERB /PHY/2023488).  
\end{acknowledgments}


%

\end{document}